\documentclass[aps,pra,twocolumn,reprint,groupedaddress]{revtex4-1}
\usepackage{physics}
\usepackage{graphicx, color, graphpap}
\usepackage[caption=false,subrefformat=parens,labelformat=simple]{subfig}
\usepackage{amsmath}
\usepackage{amsfonts}
\usepackage{lipsum}
\usepackage{amsthm}
\usepackage{tcolorbox}
\usepackage{enumitem}
\usepackage{dsfont}
\DeclareMathOperator{\spn}{span}

\DeclareMathOperator{\id}{id}
\usepackage[hyperindex]{hyperref}
\usepackage{breakurl}
\makeatletter
\newtheoremstyle{mytheorem}
  {0pt}
  {0pt}
  {}
  {10pt}
  {\itshape}
  {.---}
  {0em}
  {\thmname{#1}\thmnumber{\@ifnotempty{#1}{ }#2}%
   \thmnote{ {\the\thm@notefont(#3)}}}
\makeatother
\theoremstyle{mytheorem}

\newtheorem{lemma}{\textit{Lemma}}

\newenvironment{aeq}{\begin{equation}
\begin{aligned}
}{
\end{aligned}
\end{equation}}

\begin{document}

\title{Asymptotic Security Analysis of Discrete-Modulated Continuous-Variable \\ Quantum Key Distribution}



\author{Jie Lin}
\affiliation{Institute for Quantum Computing and Department of Physics and Astronomy, University of Waterloo, Waterloo, Ontario, Canada N2L 3G1} 
\author{Twesh Upadhyaya}
\affiliation{Institute for Quantum Computing and Department of Physics and Astronomy, University of Waterloo, Waterloo, Ontario, Canada N2L 3G1} 
\author{Norbert L\"utkenhaus}
\affiliation{Institute for Quantum Computing and Department of Physics and Astronomy, University of Waterloo, Waterloo, Ontario, Canada N2L 3G1}

\date{\today}

\begin{abstract}
Continuous-variable quantum key distribution (CV QKD) protocols with discrete modulation are interesting due to their experimental simplicity and their great potential for massive deployment in the quantum-secured networks, but their security analysis is less advanced than that of Gaussian modulation schemes. In this work, we apply a numerical method to analyze the security of discrete-modulation protocols against collective attacks in the asymptotic limit, paving the way for a full security proof with finite-size effects.  While our method is general for discrete-modulation schemes, we focus on two variants of the CV QKD protocol with quaternary modulation. Interestingly, thanks to the tightness of our proof method, we show that this protocol is capable of achieving much higher key rates over significantly longer distances with experimentally feasible parameters compared with previous security proofs of binary and ternary modulation schemes and also yielding key rates comparable to Gaussian modulation schemes. Furthermore, as our security analysis method is versatile, it allows us to evaluate variations of the discrete-modulated protocols, including direct and reverse reconciliation, and postselection strategies. In particular, we demonstrate that postselection of data in combination with reverse reconciliation can improve the key rates. 
\end{abstract}


\maketitle


\section{Introduction}

Quantum key distribution (QKD) \cite{Bennett1984, Ekert1991} is an important cryptographic primitive in the era of quantum technology since it enables two honest parties, traditionally known as Alice and Bob, to establish information-theoretically secure keys against any eavesdropper (Eve) who is bound by the laws of quantum mechanics. By now, there are plenty of QKD protocols (see Ref. \cite{Scarani2009} for a review), which can be categorized into two families according to their detection technology: discrete variable (DV) and continuous variable (CV). DV QKD protocols like the Bennett-Brassard 1984 (BB84) protocol \cite{Bennett1984} are realized by encoding the information into qubitlike degrees of freedom of photons, such as polarization and time bin, and by measuring with single-photon detectors. DV QKD enjoys great success in experimental implementations and corresponding security analyses, and can currently reach longer distances than CV QKD. However, CV QKD (e.g., see Refs. \cite{Grosshans2002, Grosshans2003a, Weedbrook2004}) uses detection technology that is widely used in modern optical (classical) communication methods, which turns those classical methods and the CV QKD apparatus into nearly identical devices. This technological similarity gives CV QKD a competitive edge for large-scale deployment in quantum-secured networks.

A main security proof technique for CV QKD is the optimality of Gaussian attacks \cite{Navascues2006, Garcia-Patron2006} for protocols with Gaussian modulation. In fact, security proofs are quite mature for CV QKD with Gaussian modulation (see \cite{Diamanti2015} for a review). However, this type of protocol puts a lot of demands on the modulation devices and classical error-correction protocols. In addition, the effect of a finite constellation needs to be taken into account carefully \cite{Jouguet2012, Kaur2019}. In probing quantumness of devices using coherent states, we notice that even a small number of coherent states have the same quantumness verification power as a Gaussian modulation of states \cite{Haseler2010}. We thus expect that a discrete-modulated CV QKD protocol will approach the performance of Gaussian-modulated CV QKD with just a few different modulation amplitudes.  However, the corresponding security proof is more involved due to missing analytical tools. The binary \cite{Zhao2009} and ternary modulation schemes \cite{Bradler2018} have been proved secure against collective attacks. Unfortunately, the key rates obtained are not tight, and the proof technique is not expected to be generalizable to discrete-modulation schemes with more states. For the quaternary modulation scheme, also known as the quadrature phase-shift keying scheme, its security was previously analyzed under the assumption of linear bosonic channels \cite{Leverrier2009} or Gaussian attacks \cite{Hirano2017, Papanastasiou2018}, which restricts Eve's ability. (Unfortunately, the analysis \cite{Leverrier2009} with the additional linear channel assumption is not expected to be tight and, thus, cannot be used as an upper bound of the key rate.)

In this work, we apply a versatile numerical method to study the security of discrete-modulated schemes with a focus on the quaternary modulation scheme. Specifically, we analyze two variants of the quadrature phase-shift keying modulation scheme: one with homodyne detection and the other with heterodyne detection. Our method enables us to obtain tight key rates against collective attacks in the asymptotic limit.  Recently, we noticed an independent work \cite{Ghorai2019} that analyzes the asymptotic security of the quaternary modulation scheme with heterodyne detection. In this security analysis \cite{Ghorai2019}, Ghorai \textit{et al.} use a reduction to the Gaussian optimality proof method and apply a semidefinite program (SDP) technique with a photon-number cutoff assumption. We emphasize here that our proof technique is quite different from their work, and, in particular, we do not invoke the arguments of Gaussian optimality. For this reason, we also directly compare our results with their results in this work. Remarkably, compared with the similar heterodyne scheme considered in Ref. \cite{Ghorai2019}, we obtain quite higher key rates. Furthermore, our approach can be extended to variants of the protocol using homodyne measurements. Since our method does not rely on the arguments of Gaussian optimality, it also allows us to investigate the effects of the postselection of data \cite{Silberhorn2002, Heid2007}, which is not considered in Ref. \cite{Ghorai2019} due to their proof technique. Postselection is commonly done for the classical telecommunication protocols and DV QKD protocols to discard noisy data and to improve the performance of the protocols. However, postselection strategies are currently not compatible with the Gaussian optimality proof technique, since the relevant states are non-Gaussian and Gaussian attacks are not expected to be optimal in the presence of postselection. Previously, postselection for discrete-modulation schemes was considered under a restricted class of attacks \cite{Heid2006, Hirano2017}. In this work, we show that postselection can improve the key rates under collective attacks. Finally, we remark that our security proof method works for both direct reconciliation and reverse reconciliation protocols. However, we focus on reverse reconciliation in this work, since reverse reconciliation is known to have better performance than the direct reconciliation in terms of transmission distances. 

For our security analysis, we rely on the numerical key rate optimization methods developed in Refs. \cite{Coles2016, Winick2018}, and we use the version of Ref. \cite{Winick2018} to prove the security against collective attacks in the asymptotic limit. Another contribution of this work is that we further develop the framework to handle the classical postprocessing for the numerical method presented in Ref. \cite{Winick2018}. This development allows us to study the postselection strategies and also simplifies some aspects of the numerical calculation. In order to perform such an optimization numerically, we impose the photon-number cutoff assumption, which is the same assumption considered in Ref. \cite{Ghorai2019}.  Although, ultimately, one would like to prove the security without this assumption, this assumption is reasonable because we numerically verify that our key rate results do not depend on the choice of cutoff when the cutoff photon number is much larger than the mean photon number of each received state. We leave it as future work to remove this assumption. It is also interesting to point out that even though we demonstrate our proof method on only the quaternary modulation scheme here, our approach can be easily generalized to other discrete-modulation schemes beyond four coherent states.

The rest of the paper is outlined as follows. In Sec. \ref{sec:description}, we present two variants of the protocol: Protocol 1 uses homodyne detection and uses only two out of four states to generate keys; protocol 2 uses heterodyne detection and encodes two-bit information in each round. In Sec. \ref{sec:security}, we first review the relevant numerical approach used for this work, discuss the photon-number cutoff assumption, and then present the specific setup of the optimization problems for those two protocols such as choices of constraints, the postprocessing map, and the pinching map related to the key map. We then perform simulations and show the simulation results in Sec. \ref{sec:simulation}.  Finally, we summarize the results and provide insights for future directions in Sec. \ref{sec:outlook}. We discuss some technical details in the Appendixes. Particularly, we present a complete framework for postprocessing in Appendix \ref{app:postprocessing}.

\section{Description of protocols}\label{sec:description}
In the following description, let $[N]$ denote the set of positive integers from 1 to $N$. A coherent state with an amplitude $\alpha$ or $\gamma$ is denoted by $\ket{\alpha}$ or $\ket{\gamma}$.
\subsection{Protocol 1 (homodyne detection)}

\begin{enumerate}

\item[(1).] \textit{State preparation}.---For each round $k \in [N]$ (where $N$ is sufficiently large),  according to the probability distribution $(\frac{p_A}{2},\frac{p_A}{2},\frac{1-p_A}{2},\frac{1-p_A}{2})$, Alice prepares a coherent state $\ket{\psi_k}$ from the set $\{\ket{\alpha}, \ket{-\alpha}, \ket{i\alpha}, \ket{-i\alpha}\},$ where $\alpha \in \mathbb{R}$ is predetermined. Alice sends this state to Bob through an insecure quantum channel. 
\item[(2).] \textit{Measurement}.---After receiving Alice's state, Bob performs a homodyne measurement on the state. Bob first generates a random bit $b_k $ according to the probability distribution $(p_B, 1-p_B)$. If $b_k = 0$, he measures the $q$ quadrature and if $b_k=1$, he measures the $p$ quadrature. He obtains the measurement outcome $y_k \in \mathbb{R}$. 
\item[(3).] \textit{Announcement and sifting}.---After $N$ rounds of first two steps, Alice and Bob communicate via the authenticated classical channel to partition all the rounds $[N]$ into four subsets defined as
\begin{aeq}
 \mathcal{I}_{qq} &= \{k \in [N]:  \ket{\psi_k}  \in \{\ket{\alpha}, \ket{-\alpha}\}, b_k = 0\},\\
 \mathcal{I}_{qp} &= \{k \in [N]:  \ket{\psi_k}  \in \{\ket{\alpha}, \ket{-\alpha}\}, b_k = 1\},\\
 \mathcal{I}_{pq} &= \{k \in [N]: \ket{\psi_k}  \in \{\ket{i\alpha}, \ket{-i\alpha}\}, b_k =0\},\\
 \mathcal{I}_{pp} &= \{k \in [N]:  \ket{\psi_k}  \in \{\ket{i\alpha}, \ket{-i\alpha}\}, b_k = 1\}.\\
\end{aeq}Then Alice and Bob randomly select a small test subset $\mathcal{I}_{qq, \text{test}} \subset \mathcal{I}_{qq}$. This selection allows them to define $\mathcal{I}_{\text{key}}$ as the subset of $\mathcal{I}_{qq}$ after removing $\mathcal{I}_{qq,\text{test}}$ and to define $\mathcal{I}_{\text{test}} =  \mathcal{I}_{qq,\text{test}} \cup \mathcal{I}_{qp} \cup \mathcal{I}_{pq} \cup \mathcal{I}_{pp}.$ Let $m$ denote the size of the index set $\mathcal{I}_{\text{key}}$ and let $f$ be a bijective function from $[m] =\{1, 2, \dots, m\}$ to $\mathcal{I}_{\text{key}}$.  After sifting, Alice sets her string $\mathbf{X}=(x_1, x_2, \dots, x_m)$ according to the rule
\begin{equation}
\forall j \in [m], \; x_{j}=\begin{cases}
0 & \text{if } \ket{\psi_{f(j)}} = \ket{\alpha}, \\
1 & \text{if } \ket{\psi_{f(j)}} = \ket{-\alpha}.\\
\end{cases}
\end{equation}

\item[(4).] \textit{Parameter estimation}.---Alice and Bob perform parameter estimation by disclosing all the information in the rounds indexed by the test set $\mathcal{I}_{\text{test}} $. To perform such an analysis, they process the data by computing quantities like the first and second moments of $q$ and $p$ quadratures conditioned on each of four states that Alice sends. These quantities allow them to constrain their joint state $\rho_{AB}$. They then calculate the secret key rate according to the optimization problem in Eq. (\ref{eq:optimization}). If their analysis shows that no secret keys can be generated, then they abort the protocol. Otherwise, they proceed.
\item[(5).] \textit{Reverse reconciliation key map}.---Bob performs a key map to obtain his raw key string. This key map discretizes his measurement outcome $y_k$ to an element in the set $\{0,1,\perp\}$ for each $k \in \mathcal{I}_{\text{key}}$.  For each $j \in [m]$, Bob sets $z_{j}$ according to the rule
\begin{equation}
z_{j}=\begin{cases}
0 & \text{if }y_{f(j)} \in [\Delta_c, \infty), \\
1 & \text{if }y_{f(j)} \in (-\infty, -\Delta_c], \\
\perp & \text{if } y_{f(j)} \in (-\Delta_c, \Delta_c).
\end{cases}
\end{equation}

Note that $\Delta_c \geq 0$ is a parameter related to the postselection of data. A protocol without postselection can set $\Delta_c =0$. At the end of this process, Bob has a string $\mathbf{Z} = (z_1, z_2, \dots, z_m).$ In communication between Alice and Bob, positions with the symbol $\perp$ are deleted from their strings. With a slight abuse of notation, we use $\mathbf{X}, \mathbf{Z}$ to mean the strings after removing the positions related to $\perp$. $\mathbf{Z}$ is called the raw key string. 
\item[(6).] \textit{Error correction and privacy amplification}.---Bob chooses a suitable error-correction protocol and a suitable privacy-amplification protocol according to the security analysis done in the parameter estimation step and communicates the choices to Alice. Alice and Bob then apply the chosen error-correction protocol and privacy-amplification protocol to generate a secret key. 
\end{enumerate}

We remark on the asymmetric roles of these four states and asymmetric choices of quadrature measurements considered here. In this specific setup, Alice and Bob use only signal states $\{\ket{\alpha}, \ket{-\alpha}\}$ and $q$ quadrature measurement data to generate keys and use all other combinations to probe eavesdropping activities. In the asymptotic limit, we can set $p_A$ and $p_B$ arbitrarily close to 1 so that the sifting factor of the protocol is 1 (in the absence of postselection) \cite{Lo2005}. However, for a finite number $N$, it is unlikely that $p_A$ and $p_B$ can be arbitrarily close to 1, since one needs to balance the trade-off between the sifting factor and the accuracy of parameter estimation. In this case, one needs to optimize the choices of  $p_A$ and $p_B$ for a given choice of $N$. The reason that we choose this asymmetric version here is to simplify some numerical calculation and to maximize the sifting factor. We may also consider another variant of this protocol, that is, allowing Alice and Bob to generate keys from both $\mathcal{I}_{qq}$ and $\mathcal{I}_{pp}$. Then for $p_A = p_B =\frac{1}{2}$, the protocol has $\frac{1}{2}$ sifting factor (in the absence of postselection). However, we point out that the essential idea of our security proof in the asymptotic limit is the same for these different variations.

\subsection{Protocol 2 (heterodyne detection)}

\begin{figure}
\includegraphics[width=0.9\linewidth]{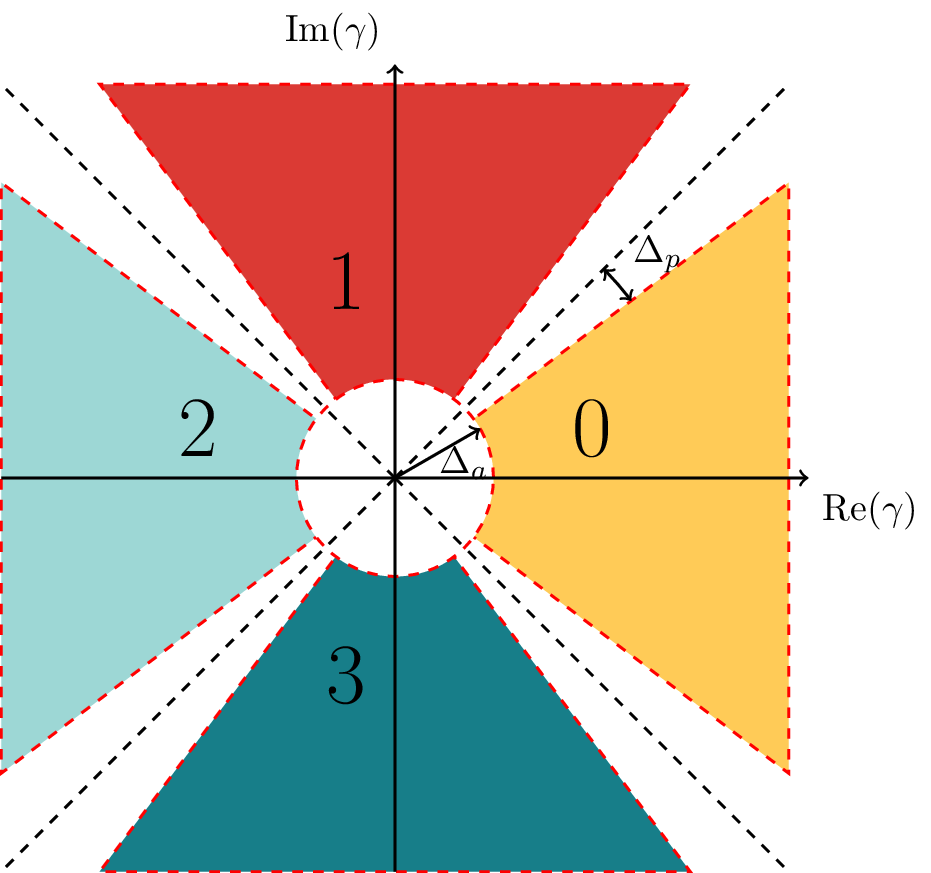}
\caption{\label{fig:keymap}Key map for protocol 2. When Bob has a measurement outcome $\gamma \in \mathbb{C}$, if $\gamma$ is in one of the four shaded areas, then Bob maps the measurement outcome to the corresponding value of that area for his key string. If $\gamma$ is not in the shaded areas, Bob obtains the symbol $\perp$. $\Delta_a$ and $\Delta_p$ are two parameters related to postselection.}
\end{figure}

This variant differs from protocol 1 in steps 2, 3, and 5. 
\begin{itemize}[leftmargin=*]
\item[(1).] \textit{State preparation}.---Like protocol 1, Alice prepares one of those four signal states with an equal probability ($p_A = \frac{1}{2}$) and sends to Bob. 
\item[(2').] \textit{Measurement}.---Upon receiving Alice's state, Bob performs a heterodyne measurement on the state, which can be described by a positive operator-valued measure (POVM) $\{E_{\gamma} = \frac{1}{\pi}\dyad{\gamma}{\gamma}: \gamma \in \mathbb{C}\}$. After applying this POVM, he obtains the measurement outcome $y_k  \in \mathbb{C}$. 
\item[(3').] \textit{Announcement and sifting}.---After $N$ rounds of first two steps, Alice and Bob determine a small subset $\mathcal{I}_{\text{test}} \subset [N]$. Rounds indexed by the set $\mathcal{I}_{\text{test}}$ are used for parameter estimation. They use the remaining rounds indexed by $\mathcal{I}_{\text{key}} = [N] /\mathcal{I}_{\text{test}}$ to generate keys.   Let $m$ denote the size of the index set $ \mathcal{I}_{\text{key}}$ and let $f$ be a bijective function from  $[m]$ to $\mathcal{I}_{\text{key}}$. After sifting, Alice obtains her string $\mathbf{X} =(x_1, \dots, x_{m})$ by the following rule:
\begin{equation}
\forall j \in [m],\; x_{j}=\begin{cases}
0 & \text{if } \ket{\psi_{f(j)}} = \ket{\alpha},\\
1 & \text{if } \ket{\psi_{f(j)}} = \ket{i\alpha},\\
2 & \text{if } \ket{\psi_{f(j)}} = \ket{-\alpha},\\
3 & \text{if } \ket{\psi_{f(j)}} = \ket{-i\alpha}.
\end{cases}
\end{equation}
\item[(4).] \textit{Parameter estimation}.---As with protocol 1, Alice and Bob perform parameter estimation to decide whether they abort the protocol.

\item[(5').] \textit{Reverse reconciliation key map}.---Bob performs a key map to obtain his raw key string. This key map discretizes his measurement outcome $y_k$ to an element in the set $\{0,1,2, 3, \perp\}$ for each $k \in \mathcal{I}_{\text{key}}$. As $y_k \in \mathbb{C}$, we write $y_k = \abs{y_k} e^{i \theta_k}$, where $\theta_k \in [-\frac{\pi}{4},\frac{7\pi}{4})$. Bob sets each $z_{j}$ of his key string $\mathbf{Z} = (z_1, \dots, z_{m})$ according to the rule
\begin{equation}
z_{j}=\begin{cases}
0 & \text{if }\theta_{f(j)} \in[-\frac{\pi}{4} + \Delta_p, \frac{\pi}{4} - \Delta_p)  \text{ and } \abs{y_{f(j)}} \geq \Delta_a,\\
1 & \text{if }\theta_{f(j)} \in[\frac{\pi}{4} + \Delta_p, \frac{3\pi}{4} - \Delta_p) \text{ and }\abs{y_{f(j)}} \geq \Delta_a,\\
2 & \text{if }\theta_{f(j)} \in[\frac{3\pi}{4} + \Delta_p, \frac{5\pi}{4} - \Delta_p)\text{ and } \abs{y_{f(j)}} \geq \Delta_a,\\
3 & \text{if }\theta_{f(j)} \in[\frac{5\pi}{4} + \Delta_p, \frac{7\pi}{4} - \Delta_p) \text{ and }\abs{y_{f(j)}} \geq \Delta_a,\\
\perp & \text{if } \theta_{f(j)} \text{ and } \abs{y_{f(j)}} \text{ are none of the above}.
\end{cases}
\end{equation}
$\Delta_a \geq 0$ and $\Delta_p \geq 0$ are two parameters related to postselection. A protocol without postselection can set $\Delta_a= \Delta_p=0$. This key map is depicted in Fig. \ref{fig:keymap}. Like protocol 1, positions with the symbol $\perp$ are deleted from their strings. Again, we use $\mathbf{X}, \mathbf{Z}$ to mean the strings after removing the positions related to $\perp$.  $\mathbf{Z}$ is called the raw key string. 

\item[(6).] \textit{Error correction and privacy amplification}.---As with the protocol 1, they perform error correction and privacy amplification to generate a secret key.
\end{itemize}
Alice and Bob may decide to recast their strings to binary strings before or during the error-correction step depending on their choice of error-correcting code. For the consistency of our presentation, we use the alphabet $\{0,1,2,3\}$ in the following discussion.

\section{Security proof approach}\label{sec:security}
Our security proof applies the numerical key rate calculation framework developed in Refs. \cite{Coles2016, Winick2018}, which allows us to calculate the secret key rate against collective attacks in the asymptotic limit. Specifically, we implement the approach in Ref. \cite{Winick2018} to solve the key rate optimization problem in this work. We begin with reviewing relevant components of the key rate calculation. For the purpose of reviewing, we keep this part of the discussion general. We direct readers to Refs. \cite{Coles2016, Winick2018} for the derivation of the key rate optimization problem and specifically to both Ref. \cite{Winick2018} and Appendix \ref{app:postprocessing} for the technical details regarding the framework of handling the postprocessing steps of the protocol. We then discuss the photon-number cutoff assumption and possible ways to remove it. Finally, we present the specific numerical optimization problems for these two protocols considered in this work.  

\subsection{Numerical method background}\label{sec:numerical_framework}
We first review the source-replacement scheme, which allows us to recast a prepare-and-measure protocol into an entanglement-based protocol. In the asymptotic limit and under collective attacks, the key rate formula is given by the well-known Devetak-Winter formula \cite{Devetak2005}. We then briefly discuss how to reformulate the Devetak-Winter formula to obtain the relevant convex objective function for the numerical optimization. Finally, we discuss the feasible set of our optimization problem. 

\subsubsection{Source-replacement scheme}
Both protocols presented in Sec. \ref{sec:description} are prepare-and-measure schemes. When we prove the security of a prepare-and-measure scheme, we apply the source-replacement scheme \cite{Bennett1992, Grosshans2003b, Curty2004, Ferenczi2012} to obtain an equivalent entanglement-based scheme and prove the security of the entanglement-based scheme, which is easier to analyze. (This idea of equivalence is first realized in Ref. \cite{Bennett1992} for BB84 and in Ref. \cite{Grosshans2003b} for Gaussian-modulated CV protocols. References \cite{Curty2004, Ferenczi2012} identify and formulate the general principle that applies to any protocol.) The key rate that we obtain from this entanglement-based scheme is also the key rate for the corresponding prepare-and-measure scheme.

If Alice prepares states from the ensemble $\{\ket{\varphi_x}, p_x\}$ in the prepare-and-measure scheme, by the source-replacement scheme, she effectively prepares the following bipartite state in the entanglement-based scheme:
\begin{equation}
\ket{\Psi}_{AA'} = \sum_{x} \sqrt{p_x} \ket{x}_{A} \ket{\varphi_x }_{A'}, 
\end{equation}
where Alice keeps the register $A$ and sends $A'$ to Bob via a quantum channel. For both protocols considered in this work, $\{\ket{\varphi_x}\} = \{\ket{\alpha}, \ket{-\alpha}, \ket{i \alpha}, \ket{-i \alpha}\}$. To determine which state she sends to Bob, Alice performs a local measurement described by a POVM $M_A=\{M_A^x=\dyad{x}{x}\}$ on register $A$. Upon obtaining a measurement result $x$, she effectively sends to Bob the state $\ket{\varphi_x}$. Their bipartite state $\rho_{AB}$ after transmitting the register $A'$ via a quantum channel, which is described by a completely positive and trace-preserving (CPTP) map $\mathcal{E}_{A'\rightarrow B}$, is
\begin{aeq}
\rho_{AB} = (\id_A  \otimes \mathcal{E}_{A'\rightarrow B}) (\dyad{\Psi}{\Psi}_{AA'}),
\end{aeq}where $\id_A$ is the identity channel on the register $A$. Bob then performs a measurement on the register $B$ to obtain his measurement result. When Alice performs a projective measurement $\dyad{x}{x}$, the corresponding conditional state $\rho_B^{x}$ that Bob receives is defined as
\begin{equation}\label{eq:conditional_state_B}
\rho_B^{x} = \frac{1}{p_x} \Tr_A[\rho_{AB} (\dyad{x}{x}_A \otimes \mathds{1}_B)].
\end{equation}

\subsubsection{Key rate formula}

The secret key rate under collective attacks in the asymptotic limit is given by the well-known Devetak-Winter formula \cite{Devetak2005}. In the case of reverse reconciliation \cite{Grosshans2003a, Grosshans2003b}, this formula reads 
\begin{equation}\label{eq:devetak-winter}
R^{\infty}= p_{\text{pass}}\Big[I(\mathbf{X};\mathbf{Z}) -\max_{\rho \in \mathbf{S}} \chi (\mathbf{Z}{\,:\,}E)\Big],
\end{equation}
where $I(\mathbf{X};\mathbf{Z})$ is the classical mutual information between Alice's string $\mathbf{X}$ and the raw key string $\mathbf{Z}$, $\chi(\mathbf{Z}{\,:\,}E)$ is the Holevo information that quantifies Eve's knowledge about the raw key string $\mathbf{Z}$, and $p_{\text{pass}}$ is the sifting probability, that is, the probability that a given round is used for key generation after the sifting step. The set $\mathbf{S}$ contains all density operators compatible with experimental observations, which we discuss later. One can rewrite the Devetak-Winter formula as 
\begin{equation}\label{eq:devetak-winter2}
R^{\infty}=p_{\text{pass}} \Big[ \min_{\rho \in \mathbf{S}} H(\mathbf{Z}|E) -H(\mathbf{Z}|\mathbf{X})\Big],
\end{equation} 
where $H(\mathbf{Z}|E) $ and $H(\mathbf{Z}|\mathbf{X})$ are conditional von Neumann (Shannon) entropies. 

In this formula, $H(\mathbf{Z}|\mathbf{X})$ is the amount of information leakage during the error-correction step performed at the Shannon limit. In reality, since the error correction cannot be done at the Shannon limit, to take the inefficiency of error correction into account, we replace this term by the actual amount of information leakage per signal (denoted by $\delta_{\text{EC}}$) during the error-correction step. 

The crucial step to turn this problem into a convex optimization problem is to rewrite $ H(\mathbf{Z}|E)$ in terms of the bipartite quantum state $\rho_{AB}$. As shown in Refs. \cite{Coles2016, Winick2018}, the key rate expression can be reformulated as
\begin{aeq}\label{eq:keyrate_objective}
R^{\infty} = \min_{\rho_{AB} \in \mathbf{S}}D\Big(\mathcal{G}(\rho_{AB}) || \mathcal{Z}\big[\mathcal{G}(\rho_{AB})\big]\Big) - p_{\text{pass}} \delta_{\text{EC}}.
\end{aeq}In this equation, $D(\rho||\sigma) = \Tr(\rho \log_2 \rho) - \Tr(\rho \log_2 \sigma)$ is the quantum relative entropy. According to Ref. \cite{Winick2018}, $\mathcal{G}$ is a completely positive and trace nonincreasing map that describes several classical postprocessing steps of the protocol in terms of actions on the bipartite state $\rho_{AB}$. Briefly speaking, $\mathcal{G}$ is composed of an announcement map $\mathcal{A}$, a sifting projection $\Pi$, and a key map isometry $V$. The roles are explained as below.
\begin{itemize}[leftmargin=*]
\item[i)] $\mathcal{A}$ is a CPTP map that introduces classical registers $\widetilde{A}$ and $\widetilde{B}$ to store announcements and also introduces quantum registers $\overline{A}$ and $\overline{B}$ to store measurement outcomes in a coherent fashion (via isometries).
\item[ii)] The sifting projection $\Pi$ projects the state after announcements to the subspace spanned by announcement outcomes kept for the key generation purpose. 
\item[iii)] The key map isometry $V$ then utilizes classical announcement registers and quantum measurement outcome registers to perform the key map step described in the protocol and stores the result of key map to a quantum register $R$.  
\end{itemize}
Thus, $\mathcal{G}(\sigma) = V \Pi \mathcal{A}(\sigma) \Pi V^{\dagger}$ for an input state $\sigma$. We remark that this output state may be subnormalized. The normalization factor is actually $p_{\text{pass}}$. Due to this property of $\mathcal{G}$ map, the factor $p_{\text{pass}}$ is not shown in front of the first term in Eq. (\ref{eq:keyrate_objective}). Finally, $\mathcal{Z}$ is a pinching quantum channel, which completely dephases the register $R$ to read out the result of key map. If $\{Z_j\}$ is the projective measurement that can be used to obtain the result of key map from the register $R$, then, for an input state $\sigma$,
\begin{aeq}
\mathcal{Z}(\sigma) = \sum_{j} Z_j \sigma Z_j.
\end{aeq}

\subsubsection{Constraints}
We now explain the feasible set $\mathbf{S}$ of our optimization problem. The set $\mathbf{S}$ contains all bipartite density operators $\rho_{AB}$ that are compatible with experimental observations. If $\{\Gamma_i | \Gamma_i = \Gamma_i^{\dagger}, 1\leq i \leq M \}$ is the set of experimental observables for some integer $M$ and $\{\gamma_i \in \mathbb{R}  | 1 \leq i \leq M\}$ is the corresponding set of expectation values observed for each $\Gamma_i$, then the feasible set $\mathbf{S}$ of our optimization problem is
\begin{aeq}
\mathbf{S} = \{\rho_{AB} \geq 0 | \Tr(\rho_{AB} \Gamma_i) = \gamma_i, \forall i\}.
\end{aeq}In particular, we include the identity operator in the set $\{\Gamma_i\}$ to ensure $\Tr(\rho_{AB})=1$. Since Eve cannot modify Alice's system $A$ in a prepare-and-measure scheme, we additionally require that $\rho_A =\Tr_B(\rho_{AB})$ is fixed as 
\begin{aeq}
\rho_{A} = \sum_{x,x'}\sqrt{p_x p_{x'}} \bra{\varphi_{x'}}\ket{\varphi_x} \dyad{x}{x'}_A. 
\end{aeq}

A final remark is that this optimization problem is a convex optimization problem, and, in particular, it is a nonlinear SDP problem. The objective function here is the quantum relative entropy function whose arguments involve additional linear maps and it is a nonlinear convex function of $\rho_{AB}$, since quantum relative entropy is jointly convex in both arguments. The feasible set is a convex set inside the positive semidefinite cone.

\subsection{Photon-number cutoff assumption}\label{sec:cutoff_assumption}
The key rate optimization problem in Eq. (\ref{eq:keyrate_objective}) involves optimizing over all possible bipartite states $\rho_{AB}$ in the feasible set $\mathbf{S}$. The number of free variables depends on the size of $\rho_{AB}$. In order to numerically perform the optimization by computer optimization packages, we can deal only with finite-dimensional $\rho_{AB}$. In our optimization problem, as we can see from the source-replacement scheme, the dimension of Alice's system $A$ is determined by the number of different signal states that she prepares. For both protocols considered in this work, the dimension of register $A$ is 4. However, since each state that Bob receives is an optical mode and, in principle, can be manipulated by Eve, Bob' state lives in an infinite-dimensional Hilbert space $\mathcal{H}_{B}$. A basis for this Hilbert space is the photon-number states $\{\ket{n}: n\in \mathbb{N}\}$. We immediately see that Bob's POVM elements are infinite-dimensional operators and $\rho_{AB}$ is also infinite dimensional. For DV QKD, one method to reduce the dimension of the system is to apply a squashing model \cite{Beaudry2008, Tsurumaru2010, Gittsovich2014} for the protocol to obtain a lower-dimensional representation of his POVM. This reduction is possible for many DV QKD protocols, since one can explicitly formulate the squashing model. However, it is not clear how one can formulate a squashing model for CV systems. Instead, we have to impose an additional assumption in this work in order to perform the numerical optimization. This additional assumption is what we call the photon-number cutoff assumption. We impose the assumption that Bob's system lives in the Hilbert space $\mathcal{H}_{B} = \spn\{\ket{0},\ket{1},\dots, \ket{N_c}\}$ for some cutoff photon number $N_c$. Namely, if we define $\Pi_{N_c} = \displaystyle\sum_{n=0}^{N_c} \dyad{n}{n}$ with a suitable choice of photon-number cutoff parameter $N_c$ on Bob's system, we assume $\rho = \Pi_{N_c} \rho \Pi_{N_c}$ for the state $\rho$ under consideration. This assumption allows us to truncate the infinite-dimensional Hilbert space. If $N_c$ is chosen to be large enough, this assumption is a reasonable working assumption based on the following observations.

\begin{enumerate}[leftmargin=*]
\item[i)] Bob can obtain the mean photon number $n_x :=\Tr(\rho_B^x \hat{n})$ of each conditional state $\rho_B^x$ via homodyne or heterodyne measurements, where $\hat{n}$ denotes the number operator. 

\item[ii)] Since $n_x$ is known, we can pick $N_c \in \mathbb{N}$ such that $N_c $ is much larger than $n_x$ for each $x \in \{0,1,2,3\}$. For such a choice of $N_c$, the probability of finding the state to have a photon number $n \leq N_c$ is close to 1. This probability suggests that the contribution from  $ n > N_c$ photon subspace becomes negligible. Similarly, the off-diagonal blocks $(\mathds{1}-\Pi_{N_c}) \rho \Pi_{N_c}$ and $\Pi_{N_c} \rho (\mathds{1}- \Pi_{N_c})$ also have vanishing contributions. 

\item[iii)] We can increase $N_c$ to have a numerical verification that the key rate is unchanged after we choose a large enough $N_c$. 
\end{enumerate}

This photon-number cutoff assumption renders our numerical optimization of the key rate problem feasible. Although this assumption sounds reasonable as our numerical verification suggests, we emphasize that one has to deliver an exact analysis to remove this assumption for a watertight security proof. In this sense, our proof is restricted. Nevertheless, we expect the key rates of these protocols to not be affected much by this working assumption. 

We now provide some insights for removing this assumption and also for extending our current analysis to include finite-size effects and general attacks. To remove the photon-number cutoff assumption, one needs to combine our numerical optimization approach with some appropriate analytical tools. One possible approach is to develop a CV version of the squashing model. If such a squashing model exists, the key rate optimization problem then becomes a finite-dimensional problem even without the photon-number cutoff assumption. Since the effective dimension is finite, it might also be possible to apply existing tools (which are valid for finite-dimensional systems) such as the quantum de Finetti representation theorem \cite{Renner2007} or the postselection technique \cite{Christandl2009} to obtain the composable security \cite{Renner2005} of the protocol against general attacks in the finite-size regime. 

Another possible method for removing the photon-number cutoff assumption is to adopt a similar idea used in Ref. \cite{Kaur2019}, that is, using the entropy continuity bounds \cite{Shirokov2017} to provide a tight error analysis of the key rate due to the truncation of Bob's Hilbert space. If one can tightly bound the trace distance between the optimal state in the truncated subspace and the optimal state in the original infinite-dimensional space, then the existing continuity bound for the Holevo information allows us to obtain a small correction term due to the photon-number cutoff. Then, one obtains a full security proof against collective attacks in the asymptotic limit. To reach a full composable security proof along this path, one may first manage to include the finite-size effects with collective attacks and then apply appropriate tools similar to the quantum de Finetti representation theorem for CV QKD \cite{Renner2009} to take the general attacks into consideration. 

Each of these two approaches has its own challenges that need to be overcome, and, thus, we have to defer these extensions to the finite-size and coherent attacks regime without the working assumption of the photon-number cutoff to future research.

\subsection{Optimization problem for protocol 1 (homodyne detection)}

Let $\hat{a}$ and $\hat{a}^{\dagger}$ be the annihilation and creation operators of a single-mode state, respectively. They obey the commutation relation $[\hat{a}, \hat{a}^{\dagger}]=1$. To be consistent in this work, we define the quadrature operators $\hat{q}$ and $\hat{p}$, respectively, as
\begin{aeq}\label{eq:quadrature_ops}
\hat{q} &= \frac{1}{\sqrt{2}} (\hat{a}^{\dagger} + \hat{a}), \; \;
\hat{p} = \frac{i}{\sqrt{2}} (\hat{a}^{\dagger} - \hat{a}). \\
\end{aeq}They obey the commutation relation $[\hat{q},\hat{p}] = i$. 

From the homodyne measurement, we can obtain expectation values of the first and second moments of the quadrature operators $\langle \hat{q}\rangle$, $\langle \hat{q}^2\rangle$, $\langle \hat{p}\rangle$, and $\langle \hat{p}^2\rangle$. We can calculate the mean photon number of each conditional state $\rho_B^x$ from the homodyne measurement outcomes, since $\hat{n} = \frac{1}{2}( \hat{q}^2 + \hat{p}^2 - 1)=\hat{a}^{\dagger}\hat{a}.$  In addition to $\hat{n}$,  we define an operator $\hat{d} = \hat{q}^2 - \hat{p}^2 = \hat{a}^2 +  (\hat{a}^{\dagger})^2$ to utilize the second moment observations $\langle \hat{q}^2\rangle$ and $\langle \hat{p}^2\rangle$ to constrain $\rho_{AB}$.

The relevant optimization problem is
\begin{aeq}\label{eq:optimization}
\text{minimize }\; & D\big(\mathcal{G}(\rho_{AB}) || \mathcal{Z}[\mathcal{G}(\rho_{AB})]\big)\\
\text{subject to }\; & \\
& \Tr[\rho_{AB} (\dyad{x}{x}_A \otimes \hat{q})] = p_x \langle \hat{q} \rangle_x, \\
& \Tr[\rho_{AB} (\dyad{x}{x}_A \otimes \hat{p})] = p_x \langle \hat{p} \rangle_x,\\
& \Tr[\rho_{AB} (\dyad{x}{x}_A \otimes \hat{n})] = p_x \langle \hat{n} \rangle_x, \\
& \Tr[\rho_{AB} (\dyad{x}{x}_A \otimes \hat{d})] = p_x \langle \hat{d} \rangle_x, \\
& \Tr[\rho_{AB}] = 1,\\
& \Tr_B [\rho_{AB}] = \sum_{i,j=0}^3 \sqrt{p_i p_j} \bra{\varphi_j}\ket{\varphi_i} \dyad{i}{j}_{A}, \\ 
& \rho_{AB} \geq 0, 
\end{aeq}where $x \in \{0,1,2,3\}$ and $\langle \hat{q} \rangle_x, \langle \hat{p} \rangle_x, \langle \hat{n} \rangle_x$, and $ \langle\hat{d} \rangle_x$ denote the corresponding expectation values of operators $\hat{q}, \hat{p}, \hat{n}$, and $\hat{d}$ for the conditional state $\rho_B^x$, respectively. In Appendix \ref{app:truncation}, we discuss how we make these operators finite dimensional under the photon-number cutoff assumption. 

We remark that one can add more fine-grained constraints using the POVM description of homodyne measurements or using the interval operators $I_0$ and $I_1$, which we define shortly. Additional constraints can only improve the key rate, as they reduce the size of the feasible set  $\mathbf{S}$. Nevertheless, we observe that this set of constraints already gives us quite tight key rates. We expect that additional constraints will provide only marginal improvements. For the ease of presentation, we choose this set of coarse-grained constraints.

We now specify the maps $\mathcal{G}$ and $\mathcal{Z}$. For the reverse reconciliation, the postprocessing map $\mathcal{G} (\sigma)= K \sigma K^{\dagger} $ is given by the following Kraus operator: 
\begin{aeq}\label{eq:kraus_homo}
K = \sum_{z=0}^{1} \ket{z}_{R} \otimes (\dyad{0}{0}+\dyad{1}{1})_A \otimes (\sqrt{I_{z}})_B,
\end{aeq}where $I_0$ and $I_1$ are \textit{interval operators} defined in terms of projections onto (improper) eigenstates of $q$ quadrature: 
\begin{aeq}
I_0 &= \int_{\Delta_c}^{\infty} dq  \dyad{q}{q}, \; \;
I_1 = \int_{-\infty}^{{-\Delta_c}} dq  \dyad{q}{q}.\\
\end{aeq}In the definition of $K$, we project Alice's register $A$ onto the subspace spanned by the first two basis states (which are related to the states $\ket{\alpha}$ and $\ket{-\alpha}$) and act on Bob's register by interval operators from the $q$ quadrature measurement, since secret keys are generated only from the rounds where Alice sends $\ket{\alpha}$ or $\ket{-\alpha}$ and Bob performs $q$ quadrature measurements in this protocol. We remark how the postselection is handled in our security proof. Since $\Delta_c$ is a postselection parameter, the effect of postselection is reflected in the definition of interval operators which are used in the postprocessing map $\mathcal{G}$. Finally, the pinching quantum channel $\mathcal{Z}$ is described by the projections $Z_0 = \dyad{0}{0}_{R} \otimes \mathds{1}_{AB}$ and $Z_1 = \dyad{1}{1}_{R} \otimes \mathds{1}_{AB}.$

We remark that we make an additional simplification for the Kraus operator $K$. Unlike the general discussion in Sec. \ref{sec:numerical_framework} or in Ref. \cite{Winick2018}, we do not introduce the registers $\widetilde{A}, \widetilde{B}, \overline{A}$, and $\overline{B}$ in the postprocessing map $\mathcal{G}$ for this protocol. The aim of such a simplification is to reduce the total dimension of the quantum states in the key rate optimization without affecting the calculated key rates. We provide a detailed analysis in Appendix \ref{app:postprocessing} to explain why such a simplification can be made. Here, we discuss the ideas behind this simplification.
\begin{enumerate}[leftmargin=*]
\item[i)] The quantum register $\overline{A}$ is Alice's private register that stores her measurement outcome after she performs her POVM $\{M^x_A\}$ on register $A$ in a coherent fashion. Since Eve has no access to register $\overline{A}$, Alice can choose to first perform a coarse-grained measurement that introduces only the announcement register $\widetilde{A}$ and then perform a refined measurement conditioned on the announcements, which is described by a local isometry. Moreover, in the reverse reconciliation scheme, since the key map isometry $V$ does not depend on Alice's measurement outcome, the isometry for the refined measurement commutes with both the key map isometry $V$ and the pinching map $\mathcal{Z}$. As our objective function is invariant under this type of local isometries, we can choose not to apply this isometry and, thus, we do not introduce register $\overline{A}$.
\item[ii)] In the announcement step, Alice and Bob each announce whether a given round is kept for the key generation. Then the sifting process keeps only one announcement outcome, that is, when they both decide to keep the round. So, both classical registers $\widetilde{A}$ and $\widetilde{B}$ after applying the sifting projection $\Pi$ are effectively one dimensional. We then use another property of the quantum relative entropy regarding quantum-classical states to show that the calculated key rates remain the same if we omit registers $\widetilde{A}$ and $\widetilde{B}$.
\item[iii)] The key map in this protocol uses only the coarse-grained information about Bob's measurement outcomes, that is, in which interval Bob's measurement outcome lies. As with the previous discussion about register $\overline{A}$, we can view Bob's measurement in two steps. At the first step, Bob performs a coarse-grained measurement in a coherent fashion to store the desired coarse-grained outcomes in register $\overline{B}$. At the second step, Bob performs a refined measurement conditioned on the coarse-grained information to update register $\overline{B}$, which is described by a local isometry (denoted by $W$). Since the key map uses only the coarse-grained information, the key map isometry $V$ effectively needs to first undo the isometry $W$. So, we can choose not to perform the isometry $W$ and let the key map isometry $V$ use the coarse-grained information directly. The calculated key rates remain the same after we ignore the isometry $W$. In this case, the key map isometry $V$ simply copies register $\overline{B}$ to register $R$ in the standard basis. Thus, we combine these two registers and retain the name of $R$. The calculated key rates are unaffected, because copying register $\overline{B}$ to register $R$ in the standard basis is done by a local isometry, which we can omit. 
\end{enumerate}

\subsection{Optimization problem for protocol 2 (heterodyne detection)}

The optimization problem for protocol 2 has essentially the same form as described in Eq. (\ref{eq:optimization}). The differences here are that expectation values are now obtained via heterodyne detection, and the postprocessing map $\mathcal{G}$ and the pinching map $\mathcal{Z}$ have different forms, as we present shortly.  In principle, we can use additional information about second moments like $\langle \hat{q}\hat{p} \rangle$ to constrain $\rho_{AB}$ as the information becomes available via heterodyne detection. However, our calculation shows that additional constraints like this one can provide only marginal improvements on the key rates in our simulated scenarios. We expect these constraints to be more useful if we introduce squeezing in either the protocol or the simulation.   

For an input state $\rho$, heterodyne measurements give us the Husimi $Q$ function $Q(\gamma) = \frac{1}{\pi} \bra{\gamma}\rho \ket{\gamma} = \Tr(\rho E_{\gamma}),$ where $\{E_{\gamma} = \frac{1}{\pi} \dyad{\gamma}{\gamma}: \gamma \in \mathbb{C}\}$ is the POVM description of heterodyne detection. From the $Q$ function, we can also obtain values of $\langle \hat{q} \rangle$, $\langle \hat{p} \rangle$, $\langle \hat{n} \rangle$, and $\langle \hat{d} \rangle$, whose operators are functions of $\hat{a}$ and $\hat{a}^{\dagger}$, by the following equation \cite{Vogel2006}:
\begin{aeq}
\Tr[\rho \hat{f}(\hat{a},\hat{a}^{\dagger})] = \langle \hat{f}^{(A)}(\hat{a},\hat{a}^{\dagger})\rangle := \int d^2 \gamma Q(\gamma) f^{(A)}(\gamma),
\end{aeq}where $\hat{f}^{(A)} (\hat{a},\hat{a}^{\dagger})$ is the antinormally ordered operator of an operator $\hat{f}$ written in terms of $\hat{a}$ and $\hat{a}^{\dagger}$, $f^{(A)}(\gamma)$ is the corresponding expression by replacing $\hat{a}$ by $\gamma$ and $\hat{a}^{\dagger}$ by $\gamma^*$, and $d^2 \gamma = d\Re(\gamma) d\Im(\gamma).$

To write out the Kraus operator for the postprocessing map $\mathcal{G}$ including postselection, we define \textit{region operators} that tell us in which region in Fig. \ref{fig:keymap} Bob's measurement outcome lies. We express them using the polar coordinate for the integration as
\begin{aeq}\label{eq:region_operator}
R_{0} &= \frac{1}{\pi}\int_{\Delta_a}^{\infty} \int_{-\frac{\pi}{4}+\Delta_p}^{\frac{\pi}{4}-\Delta_p}  \gamma \dyad{\gamma e^{i\theta}}{\gamma e^{i\theta}} \; d\theta \; d\gamma ,\\
R_{1} &=\frac{1}{\pi} \int_{\Delta_a}^{\infty} \int_{\frac{\pi}{4}+\Delta_p}^{\frac{3\pi}{4}-\Delta_p}  \gamma \dyad{\gamma e^{i\theta}}{\gamma e^{i\theta}} \; d\theta \; d\gamma ,\\
R_{2} &= \frac{1}{\pi}\int_{\Delta_a}^{\infty} \int_{\frac{3\pi}{4}+\Delta_p}^{\frac{5\pi}{4}-\Delta_p} \gamma \dyad{\gamma e^{i\theta}}{\gamma e^{i\theta}} \; d\theta \; d\gamma ,\\
R_{3} &= \frac{1}{\pi}\int_{\Delta_a}^{\infty} \int_{\frac{5\pi}{4}+\Delta_p}^{\frac{7\pi}{4}-\Delta_p} \gamma \dyad{\gamma e^{i\theta}}{\gamma e^{i\theta}} \; d\theta \; d\gamma .\\
\end{aeq}The area of integration for each operator corresponds to the relevant region shown in Fig. \ref{fig:keymap}. Again, $\Delta_a$ and $\Delta_p$ are parameters related to postselection. 

In this case, the postprocessing map $\mathcal{G} (\sigma) = K \sigma K^{\dagger}$ is given by the Kraus operator
\begin{aeq}\label{eq:kraus_het}
K = \sum_{z=0}^3 \ket{z}_R \otimes \mathds{1}_A \otimes (\sqrt{R_{z}})_B .
\end{aeq}The pinching quantum channel $\mathcal{Z}$ is given by the projections $Z_j = \dyad{j}{j}_{R}\otimes  \mathds{1}_{AB}$  for $ j = 0, 1,2,3$, that is, for a valid input state $\sigma$: 
\begin{aeq}
\mathcal{Z}(\sigma) = \sum_{j =0}^{3} (\dyad{j}{j}_R \otimes \mathds{1}_{AB}) \sigma ( \dyad{j}{j}_{R} \otimes \mathds{1}_{AB} ).
\end{aeq}

Like protocol 1, we make a simplification for the Kraus operator $K$ by a similar line of argument.  

\subsection{Generalization to other discrete-modulation schemes beyond four coherent states}
From the description of our security proof method, we remark that this proof technique does not depend on the distribution of the statistics, whether it is Gaussian or not. Also, it is not difficult to see that our method can be generalized to analyze discrete-modulated CV QKD protocols with more coherent states. If Alice modulates using $\ell$ coherent states, then Alice's system $A$ is $\ell$ dimensional from the source-replacement scheme. In this case, the corresponding optimization problem essentially has the same form as in Eq. (\ref{eq:optimization}), except that the index $x$ now runs from $0$ to $\ell-1$ and the maps $\mathcal{G}$ and $\mathcal{Z}$ need to be modified accordingly to match the description of the protocol in a straightforward way. A guide to defining the postprocessing map $\mathcal{G}$ is also provided in Appendix \ref{app:postprocessing}.

\section{Simulation and key rates}\label{sec:simulation}
In this section, we first discuss our model for simulating experiments that execute each protocol. From the simulation, we can obtain relevant expectation values like $\langle \hat{q} \rangle$ and $\langle \hat{p} \rangle$ that we usually obtain from an actual experiment and which are the starting point of our key rate optimization problem in Eq. (\ref{eq:optimization}). Then we comment on the numerical performance of our current algorithm and discuss relevant numerical issues. Finally, we present key rates for both protocols with different variations. We emphasize that our security proof technique, of course, does not depend on the model of experiment that we use to predict the experimental behavior. 

 \subsection{Simulation model}\label{sec:simulationmodel}
To understand how the protocols behave in a realistic scenario, we simulate the quantum channel as a realistic physical channel in the absence of Eve. Such a channel in the context of optical fiber communication can be described by a phase-invariant Gaussian channel with transmittance $\eta$ and excess noise $\xi$ which is defined as
 \begin{aeq}\label{eq:excessnoise}
\xi = \frac{(\Delta  q_{\text{obs}})^2}{(\Delta q_{\text{vac}})^2} -1,
\end{aeq}where $(\Delta q_{\text{vac}})^2 $ is the variance of $q$ quadrature for the vacuum state and $(\Delta  q_{\text{obs}})^2$ is the variance of $q$ quadrature observed for the signal state. Here we consider the case where both $q$ and $p$ quadratures have the same variance. With our definition of quadrature operators, $(\Delta q_{\text{vac}})^2 =\frac{1}{2}$. In the literature, the value of excess noise is usually reported in a couple of different ways, depending on who makes the observation of $(\Delta  q_{\text{obs}})^2$. To avoid possible confusion when discussing the value of excess noise, we clarify these definitions. We use $\xi$ to mean the excess noise in the case where Alice measures $(\Delta  q_{\text{obs}})^2$ at the output of her lab and use $\delta$ in the case where Bob measures $(\Delta  q_{\text{obs}})^2$ for the received signal state. 

A natural way to simulate this phase-invariant Gaussian channel is that, when Alice prepares a coherent state $\ket{\alpha}$ and sends to Bob via this channel, the output state from the channel becomes a displaced thermal state centered at$\sqrt{\eta} \alpha$ with the variance $\frac{1}{2}(1+ \delta)$ for each quadrature. An alternative but equivalent way is that, when Alice wants to prepare a coherent state $\ket{\alpha}$, the state after preparation becomes a displaced thermal state centered at $\alpha$ with the variance $\frac{1}{2}(1+\xi)$ for each quadrature at the output of her lab. Then, the state is transmitted via a pure-loss channel, and the final output state that reaches Bob's lab is a displaced thermal state centered at $\sqrt{\eta} \alpha$ with the variance $\frac{1}{2}(1+ \eta \xi)$ for each quadrature. Therefore, we see that, for this physical channel, $\delta = \eta \xi$. In this work, we use the definition of $\xi$ when we discuss the value of excess noise. Readers should be able to translate between these two definitions by the relation $\delta = \eta \xi$.

Given a displaced thermal state centered at $\sqrt{\eta} \alpha$ with the variance $\frac{1}{2}(1+ \eta \xi)$ for each quadrature, we can then calculate our simulated values for $\langle \hat{q} \rangle$, $\langle \hat{p} \rangle$,  $\langle \hat{n} \rangle$, and $\langle \hat{d} \rangle$ (by either using quasiprobability distribution like the Wigner function or $Q$ function of the final state or expanding the final state in the photon-number basis). These values can then be supplied to the optimization problem in Eq. (\ref{eq:optimization}), which, in turn, can be solved numerically.

\subsection{About numerical algorithm and performance}\label{sec:numerical_performance}
To perform the numerical calculation of secret key rates, we apply the two-step procedure mentioned in Ref. \cite{Winick2018}. At the first step, we adopt the Frank-Wolfe algorithm \cite{Frank1956} to find a suboptimal attack that gives rise to a suboptimal bipartite state $\rho_{AB}$. At the second step, we use this suboptimal $\rho_{AB}$ to solve a linear SDP problem to obtain a reliable lower bound on the key rate, which also takes the constraint violation into consideration. The Frank-Wolfe algorithm used in the first step is an iterative first-order optimization algorithm. We start with an initial choice of $\rho_{AB}$ in the feasible set $\mathbf{S}$, and, in each iteration, we solve a linear SDP problem to update the choice of $\rho_{AB}$ until a stopping criterion is satisfied. Since this optimization algorithm may have a very slow rate of convergence near the optimal point in some scenarios, in order to have a reasonable running time, we limit the maximum number of Frank-Wolfe iterations to be 300. To solve linear SDP problems in both the first and second steps, we employ the CVX package \cite{CVX, Grant2008} and SDPT3 \cite{Toh1999, Tutuncu2003} solver in \textsc{Matlab}.

\begin{figure}[h]
\subfloat[Protocol 1]{\label{fig:numerical_performance_homo}\includegraphics[width=\linewidth]{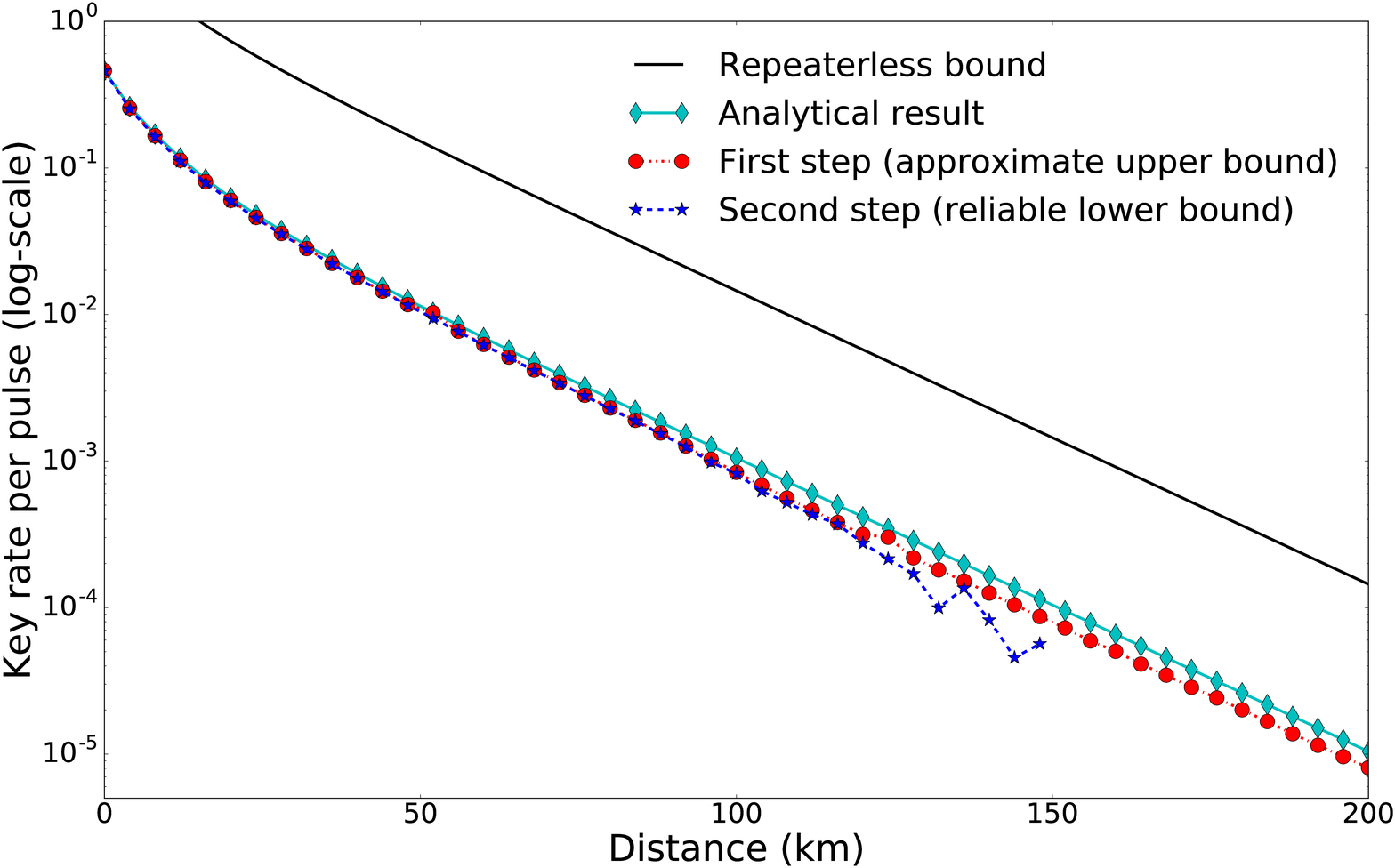}}\\
\subfloat[Protocol 2]{\label{fig:numerical_performance_het}\includegraphics[width=\linewidth]{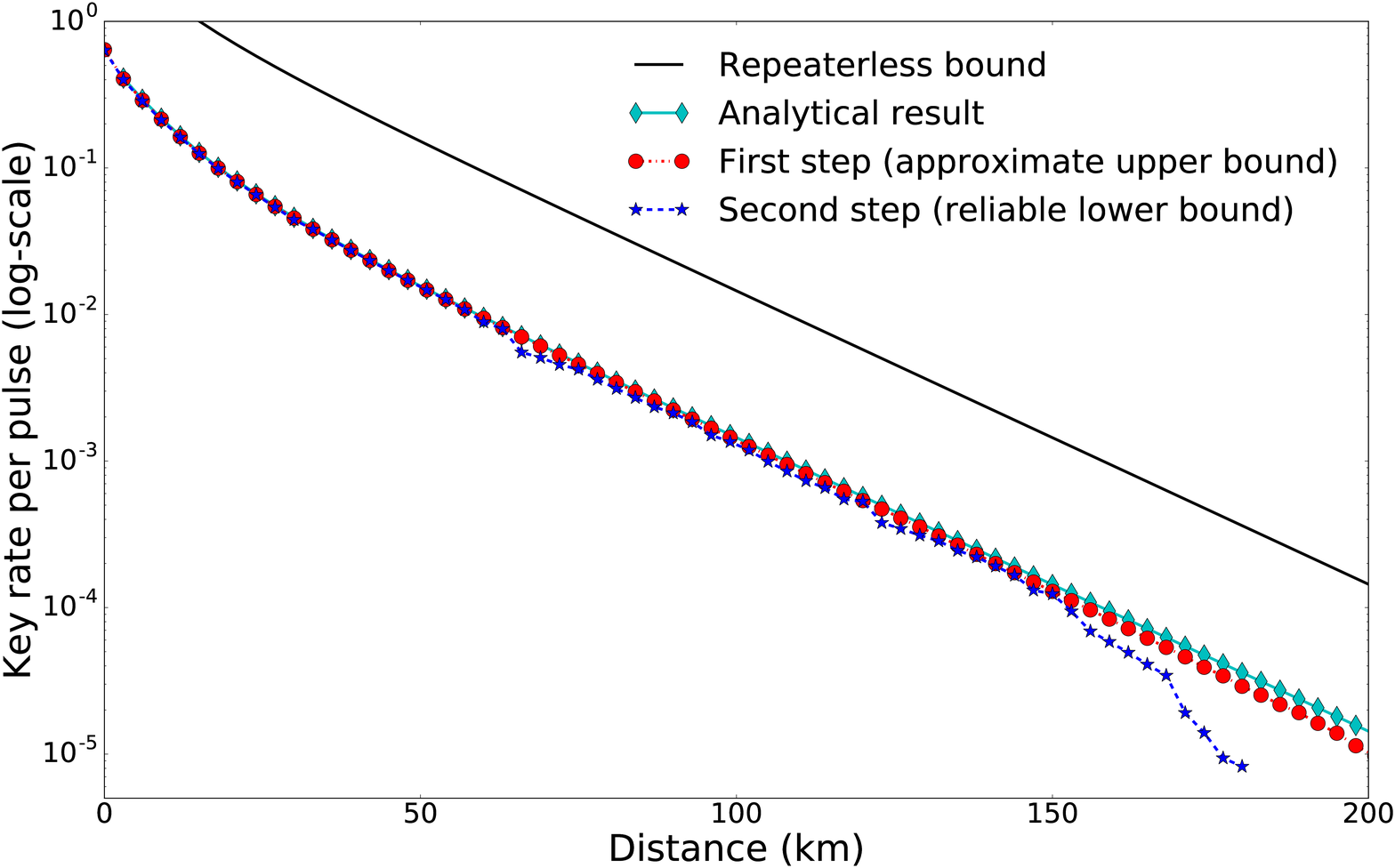}}
\caption{\label{fig:numerical_performance} Secure key rates versus the transmission distance for the pure-loss channel to demonstrate the numerical behavior of our two-step key rate calculation procedure and to compare with the analytical results (direct evaluation of the Devetak-Winter formula) for both protocols. The transmittance is $\eta = 10^{-0.02 L}$ for each distance $L$ in kilometers, and the reconciliation efficiency is $\beta = 0.95$. The curve with circle markers is the approximate upper bound from the first step, and the curve with star markers is the reliable lower bound obtained from the second step. The curve with square markers is the analytical results presented in Appendix \ref{app:lossonly}. The solid line with no markers is the repeaterless secret key capacity bound \cite{Takeoka2014,Pirandola2017}. (a) The key rate for protocol 1 (homodyne detection). The coherent state amplitude $\alpha$ is optimized via a coarse-grained search in the interval [0.36, 0.6]. (b) The key rate for protocol 2 (heterodyne detection).  $\alpha$ is optimized via a coarse-grained search in the interval [0.6, 0.95]. }
\end{figure}

In Fig. \ref{fig:numerical_performance}, we plot the results of the first step and the second step for protocols 1 and 2 in the case of the pure-loss channel ($\xi = 0$) which is discussed in detail in Sec. \ref{sec:lossonly}. Here, we utilize this figure to illustrate some aspects of the numerical analysis. The result from the first step can be treated as an approximate upper bound, since it is given by a suboptimal $\rho_{AB}$. (It is only approximate, because the feasible set $\mathbf{S}$ might be enlarged either due to the coarse-graining of constraints or due to the numerical constraint violation.) The result of the second step is a reliable lower bound on the key rate. There are three regions in each plot. Since both plots have similar numerical behaviors, we take Fig. \ref{fig:numerical_performance_homo} as an example to discuss these three regions. The first region is where both steps give essentially the same results, as we can see for points between 0 and around 120 km. The second region is between around 120 and 150 km, where there is a noticeable gap between our approximate upper bound and the reliable lower bound. This gap is an indicator that the first-step algorithm fails to find a good suboptimal $\rho_{AB}$. In fact, the chosen number of maximum iterations for the first-step Frank-Wolfe algorithm is reached for those points. The third region is beyond 150 km where the lower bound is missing in the plot. This result is because the suboptimal $\rho_{AB}$ from the first step also has some noticeable constraint violation when the first step terminates prematurely after 300 iterations. Since we take constraint violations into account in the second-step calculation to obtain a reliable lower bound according to Ref. \cite{Winick2018}, these points correspond to the case where the lower bound obtained is zero. To obtain a better lower bound, one needs to improve the result of the first-step calculation. There are several possible ways to improve the first-step result:
\begin{itemize}[leftmargin=*]
\item[i)] replacing the Frank-Wolfe algorithm by other optimization algorithms,
\item[ii)] using a different SDP solver,
\item[iii)] choosing a different initial point ($\rho_{AB}$) for the first step, and
\item[iv)] increasing the number of iterations. 
\end{itemize}

The main reason behind these alternatives is that different solvers and different algorithms can have different rates of convergence and, thus, can potentially give better results within the time limit. Since the aim of this work is not about optimizing the numerical optimization algorithm, we choose to report results based on our current choice of algorithm and solver mentioned before with the limitation of 300 iterations in the first step, and we see that such a choice works well in many scenarios.

For all the remaining figures in this work, we report only the reliable lower bound obtained from the second step. For some of the curves shown in this work, even though there are data points from the second and third scenarios mentioned above, which are not compatible with the general trend of the curve, these numbers can still be safely interpreted as reliable (but very pessimistic) secret key rates. We may expect that, if we improve the optimization algorithm (which is not the goal of this work), we can obtain smoother curves. It is also interesting to point out that, when we add some nonzero excess noise, the curves that we obtain (shown in later sections) can be smoother than the loss-only curves. We can understand this behavior from the fact that the rank of the density matrix $\rho_{AB}$ is much smaller than its dimension for the loss-only case, and, thus, the problem is not numerically well conditioned. One can improve on this aspect if one can reformulate the problem using a lower-dimensional representation. 

\subsection{Loss-only scenario: Comparison to analytical results}\label{sec:lossonly}
We first present the results for the loss-only scenario, that is, $\xi =0$. For this scenario, we can also obtain an analytical result to have a direct comparison with our numerical result. A direct evaluation of the Devetak-Winter formula is possible in this scenario since we can determine Eve's relevant conditional states (up to irrelevant unitaries). As shown in Ref. \cite{Heid2006}, in the loss-only case, we need only to consider the generalized beam-splitting attack. When Alice sends $\ket{\alpha_x}_{A'}$ to Bob, the state becomes $\ket{\sqrt{\eta}\alpha_x}_B \ket{\sqrt{1-\eta}\alpha_x}_E$ after the pure-loss channel. Eve's conditional states conditioned on Alice's string value $x$ and Bob's raw key string value $z$ effectively live in a two-dimensional subspace for protocol 1 and a four-dimensional subspace for protocol 2. This result makes the direct analytical evaluation possible. We leave the procedure of this analytical evaluation to Appendix \ref{app:lossonly}. For the numerical key rate optimization, the loss-only scenario follows as a special case of the noisy scenario (using $\xi = 0$), which we discuss in later sections. 

A pure-loss channel is characterized by its transmittance $\eta = 10^{-(\alpha_{\text{att}} L/10)}$ for each distance $L$ in kilometers with the attenuation coefficient $\alpha_{\text{att}}$, which is 0.2 dB/km for the relevant communication fiber. One may take the quantum efficiency of realistic homodyne and heterodyne detectors into account. A simple but pessimistic way to deal with the detector efficiency is that the loss due to the imperfect detector is also attributed to Eve. In such a worse-case scenario, we can define the total transmittance as $\eta = \eta_{\text{det}} 10^{-0.02 L}$, where $\eta_{\text{det}}$ is the quantum efficiency of the detectors. If one defines an effective distance $L_0$ for the detector inefficiency, that is, $\eta_{\text{det}} = 10^{-0.02 L_0}$, then $L_0$ is less than 13 km for practical homodyne and heterodyne detectors with the quantum efficiency $\geq$ 55\% \cite{Jouguet2013}. For the ease of presentation and convenience of comparison with other works using different values of detector efficiency, we set $\eta_{\text{det}} =1$ in this work unless noted otherwise. One may obtain the key rate value corresponding to a realistic value of efficiency by subtracting the effective distance $L_0$ from all relevant figures. 

We plot the key rate versus transmission distance in the loss-only scenario for protocol 1 in Fig. \ref{fig:numerical_performance_homo} and for protocol 2 in Fig. \ref{fig:numerical_performance_het}. For both protocols, we plot both the numerical key rate calculation results and the key rate that can be obtained by a direct evaluation of the Devetak-Winter formula. Interestingly, we see that our numerical results are close to the analytical results for both protocols up to a distance around 120 km. Above 120 km, we notice that there is a visible gap between our approximate upper bound and the reliable lower bound, which indicates there is room for improvement on the numerical algorithm. We also notice that our first-step result is slightly lower than the analytical result. The reason is that, by analytical analysis, we know the feasible set $\mathbf{S}$ effectively should contain only one state (up to irrelevant unitaries from the perspective of entropy evaluation). However, we use coarse-grained constraints in our numerical optimization, and, thus, the feasible set $\mathbf{S}$ is actually enlarged. We expect that, if all fine-grained constraints are used, we should be able to reproduce the analytical results in this loss-only scenario (when a better optimization algorithm is used).  

We also include the repeaterless secret key capacity bound for the pure-loss channel \cite{Takeoka2014,Pirandola2017} in both plots, that is, $R^{\infty} \leq -\log_2(1-\eta)$. With the reconciliation efficiency $\beta = 0.95$ (explained in the next section),  the key rate for protocol 1 is roughly $\frac{1}{15}$ of the secret key capacity bound, and the key rate for protocol 2 is approximately $\frac{1}{10}$ of the bound. Since Gaussian modulation schemes with the perfect reconciliation efficiency can reach $\frac{1}{2}$ of this bound \cite{Pirandola2017}, we see that the performance of the quaternary modulation scheme is not far away from that of the Gaussian modulation schemes in the loss-only scenario.

\subsection{Noisy scenario: Protocol 1}

\subsubsection{Simulated statistics and error-correction cost}
We now consider the noisy scenario with nonzero excess noise $\xi$. From the homodyne measurement, for each $\alpha_x \in \{\alpha, -\alpha ,i\alpha, -i\alpha\},$ the simulated statistics is given as
\begin{aeq}
\langle\hat{q}\rangle_x &= \sqrt{2\eta} \Re(\alpha_x),\\
\langle\hat{p}\rangle_x &=  \sqrt{2\eta} \Im(\alpha_x),\\
\langle\hat{n}\rangle_x &= \eta \abs{\alpha_x}^2 + \frac{\eta \xi}{2}, \\
\langle\hat{d}\rangle_x &=  \eta [\alpha_x^2 + (\alpha_x^{*})^{2}].\\
\end{aeq}With these values specified, we perform the optimization to bound Eve's information. 

Since we simulate the experimental behavior and the cost of error correction is not a part of the optimization, we now present the analytical formula to estimate $\delta_{\text{EC}}$  from the simulated statistics and numerically evaluate the formula. In this protocol, we use only $\ket{+\alpha}, \ket{-\alpha}$ ($\alpha \in \mathbb{R}$) and the $q$ quadrature measurement to generate keys. After Bob performs his key map, Alice and Bob effectively communicate via a binary channel for the purpose of error correction. From the simulation, the probability distributions of Bob's $q$ quadrature measurement outcomes for conditional states $\rho_B^0$ and $\rho_B^1$ are
\begin{aeq}
P(q|0) = \frac{1}{\sqrt{\pi (\eta \xi + 1)}} e^{\frac{-(q-\sqrt{2\eta} \alpha)^2}{\eta \xi+1}},\\
P(q|1) = \frac{1}{\sqrt{\pi (\eta \xi + 1)}} e^{\frac{-(q+\sqrt{2\eta} \alpha)^2}{\eta \xi+1}}.\\
\end{aeq}Since we allow postselection with the cutoff parameter $\Delta_c$, the sifting probability reads
\begin{aeq}
p_{\text{pass}} &=  1 - \frac{1}{2}\int_{-\Delta_c}^{\Delta_c} P(q|0) dq  - \frac{1}{2}\int_{-\Delta_c}^{\Delta_c} P(q|1) dq.
\end{aeq}The error probability between Alice's and Bob's strings is
\begin{aeq}
e = \frac{1}{p_{\text{pass}} }\Big(\frac{1}{2} \int_{-\infty}^{-\Delta_c} P(q|0) dq + \frac{1}{2} \int_{\Delta_c}^{\infty} P(q|1) dq\Big). 
\end{aeq}

For the error correction performed at the Shannon limit, we have $\delta_{\text{EC}} = H(\mathbf{Z}|\mathbf{X}) = h (e)$, where $h(x) = -x \log_2(x) -(1-x) \log_2(1-x)$ is the binary entropy function. To take into account the inefficiency of error correction, we first write $\delta_{\text{EC}} = H(\mathbf{Z}|\mathbf{X})= H(\mathbf{Z}) - I(\mathbf{X};\mathbf{Z})$ in terms of $I(\mathbf{X};\mathbf{Z})$ and then scale $I(\mathbf{X};\mathbf{Z})$ to be $\beta I(\mathbf{X};\mathbf{Z})$ where $\beta$ is the reconciliation efficiency whose value is usually reported in the CV QKD literature. Therefore,
\begin{aeq}\label{eq:delta_ec}
\delta_{\text{EC}} &=  H(\mathbf{Z}) - \beta I(\mathbf{X};\mathbf{Z})\\
& = (1-\beta) H(\mathbf{Z})  + \beta H(\mathbf{Z}|\mathbf{X}) \\
& = (1-\beta) H(\mathbf{Z})  + \beta h(e).
\end{aeq}

In this work, we use $\beta =0.95$ in all figures unless mentioned otherwise.

\subsubsection{Key rates for protocol 1}

\begin{figure}
\subfloat[$L= 20$ km]{\label{fig:hom_alpha_d20}\includegraphics[width=0.5\linewidth]{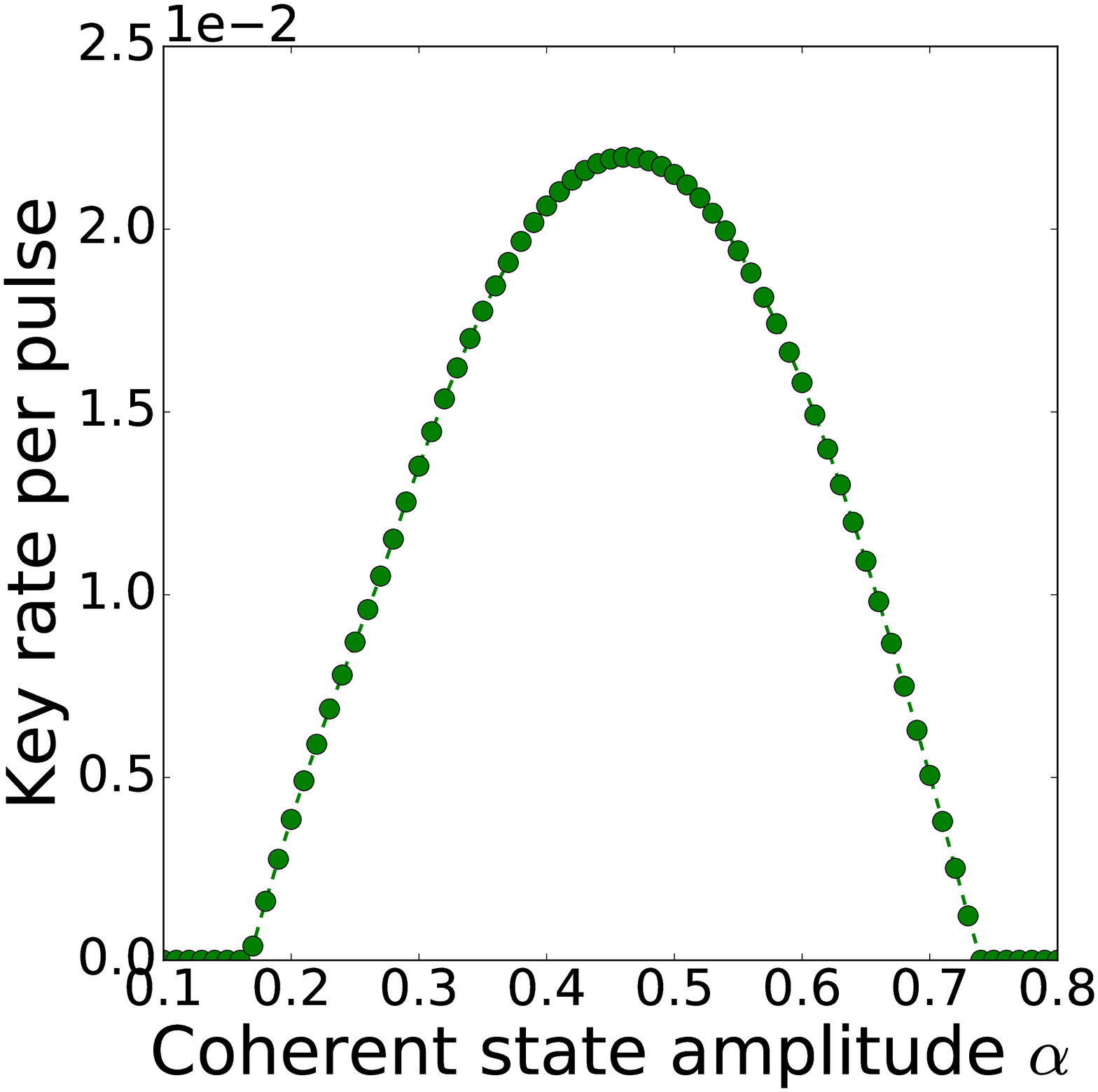}}
\subfloat[$L= 50$ km]{\label{fig:hom_alpha_d50}\includegraphics[width=0.5\linewidth]{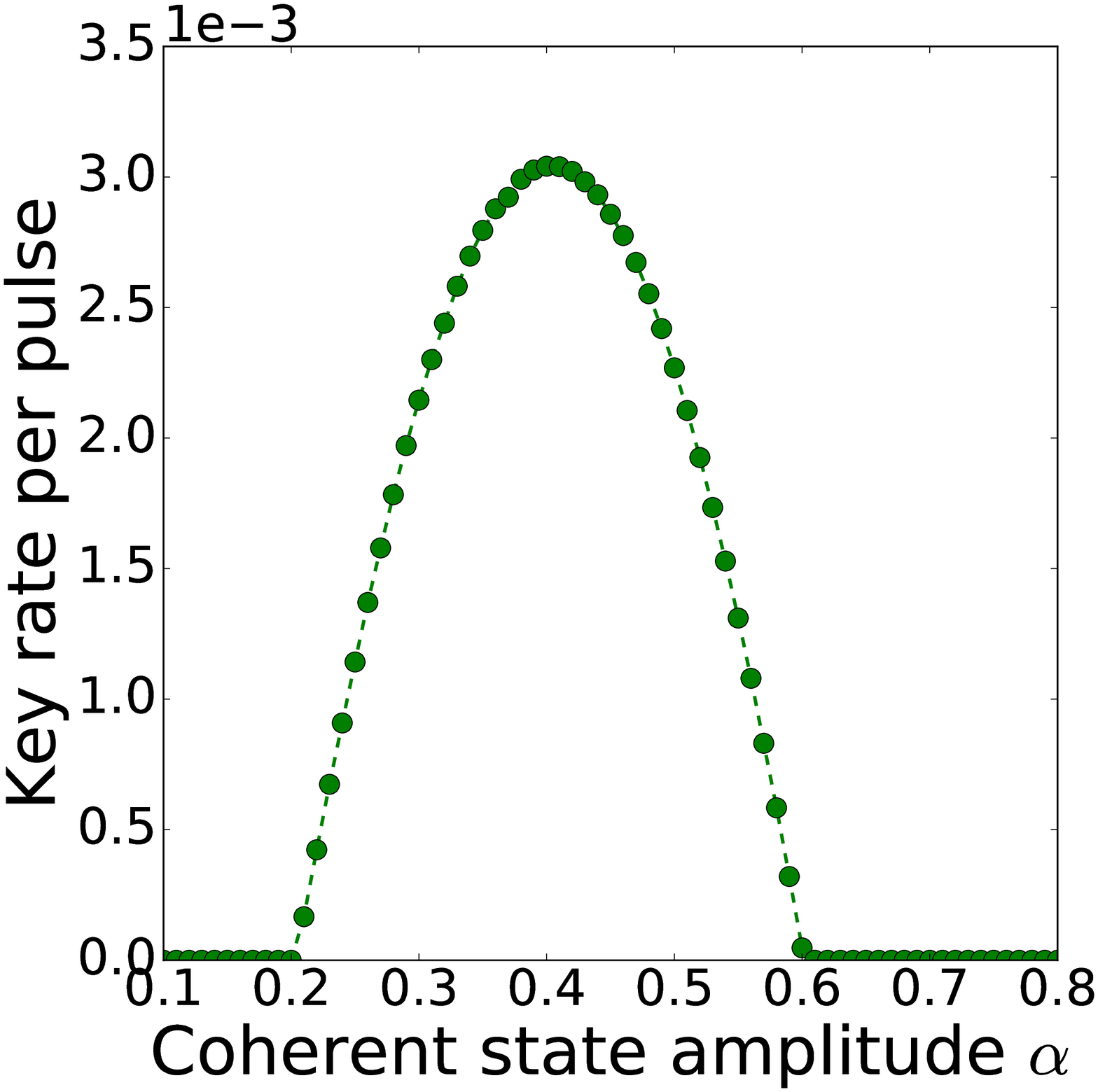}}\\
\subfloat[$L= 80$ km]{\label{fig:hom_alpha_d80}\includegraphics[width=0.5\linewidth]{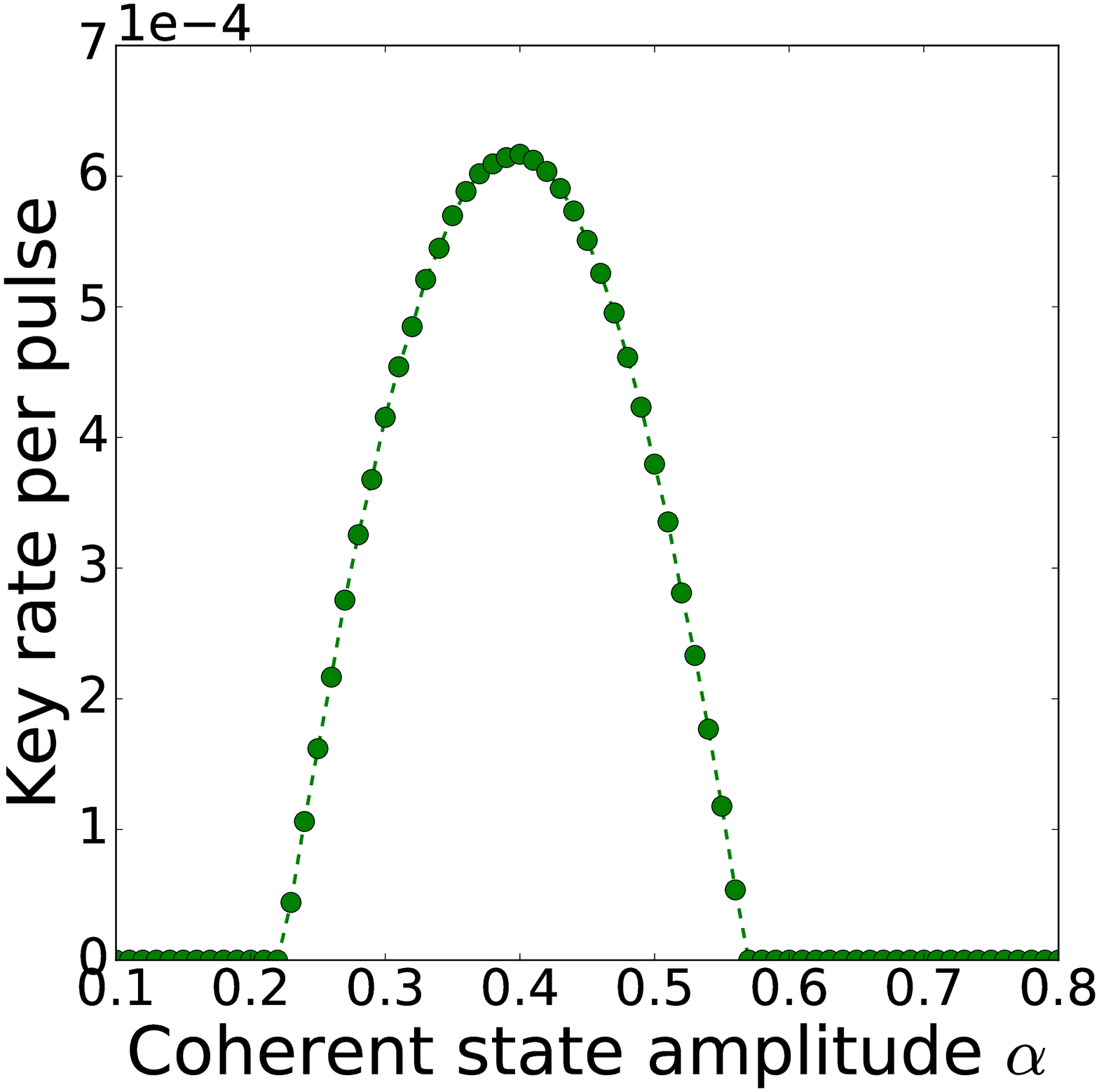}}
\subfloat[$L= 100$ km]{\label{fig:hom_alpha_d100}\includegraphics[width=0.5\linewidth]{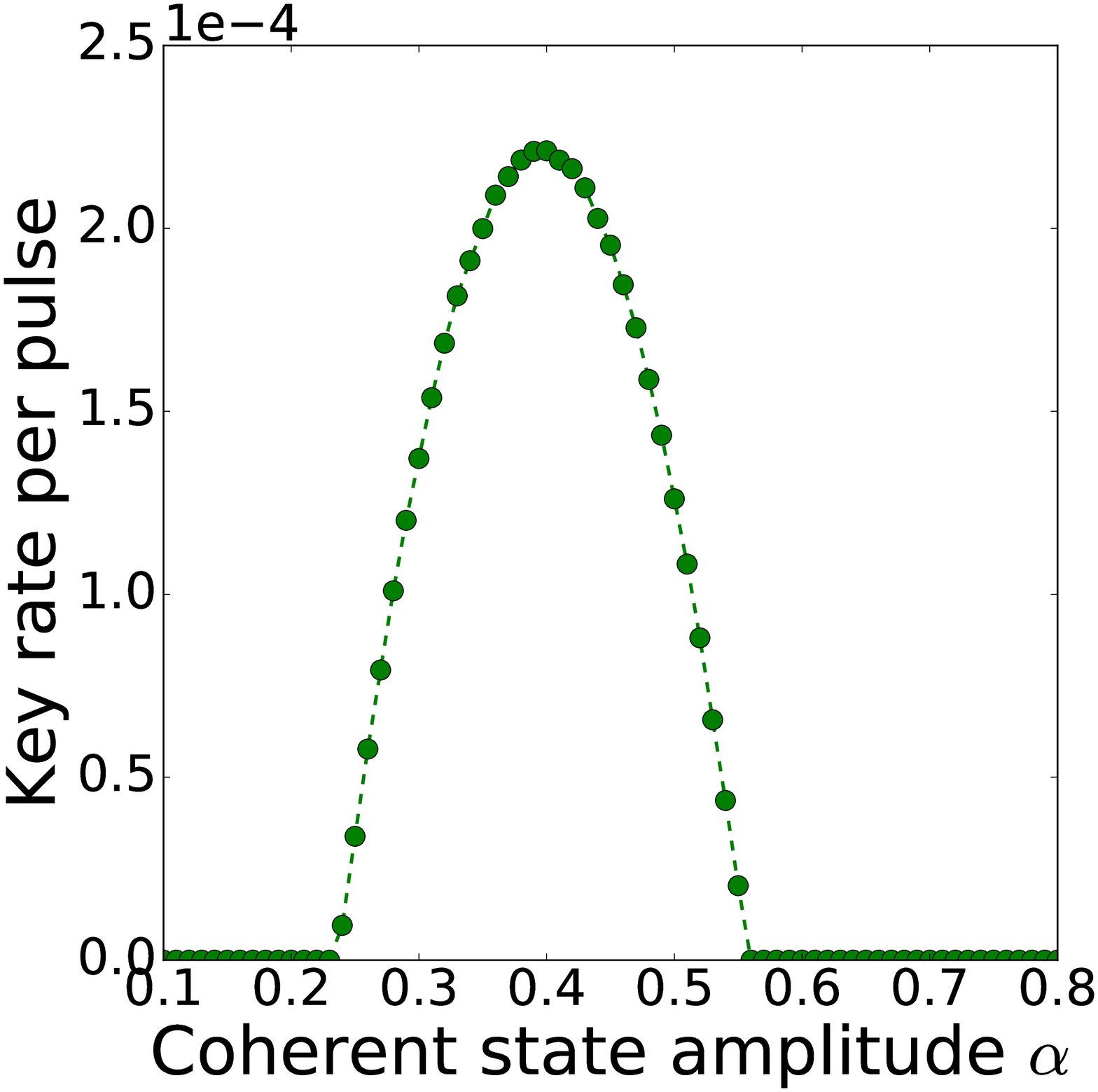}}
\caption{\label{fig:hom_alpha} Secure key rate for protocol 1 versus coherent state amplitude $\alpha$ for selected choices of distances (a) $L = 20$ km, (b) $L = 50$ km, (c) $L = 80$ km, and (d) $L = 100$ km with the excess noise  $\xi = 0.01$ and reconciliation efficiency $\beta = 0.95$. }
 \end{figure}

We first investigate the optimal choice of coherent state amplitude $\alpha$ in the absence of postselection, that is, $\Delta_c = 0$. In Fig. \ref{fig:hom_alpha}, we plot the key rate versus the choice of $\alpha$ for a selected set of distances in the case of the excess noise $\xi = 0.01$. The optimal choice of $\alpha$ for each distance $L= 20, 50, 80, 100$ km lies around 0.4, corresponding to a mean photon number of 0.16 from Alice's source. We also see that the optimal choice does not change significantly for different distances. This observation allows us to search in a restricted interval when we optimize $\alpha$ to maximize the key rate for each transmission distance.

\begin{figure}[h]
\includegraphics[width=\linewidth]{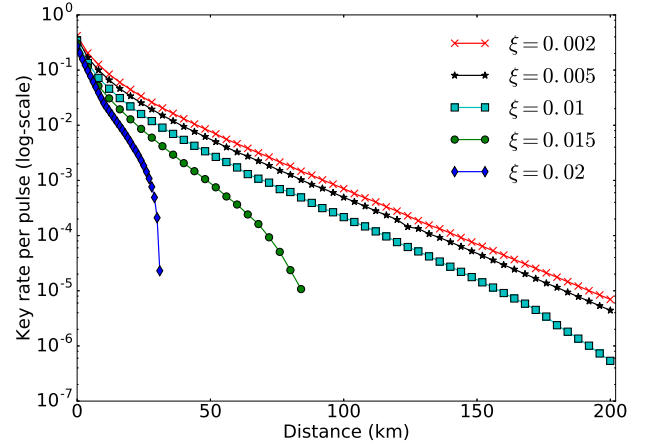}
\caption{\label{fig:hom_no_ps} Secure key rate versus the transmission distance for protocol 1 (with homodyne detection) for different values of the excess noise, from top to bottom, $\xi = 0.002, 0.005, 0.01, 0.015, 0.02$. The coherent state amplitude $\alpha$ is optimized via a coarse-grained search in the interval $[0.35, 0.6]$, the transmittance is $\eta = 10^{-0.02 L}$ for each distance $L$ in kilometers and the reconciliation efficiency is $\beta = 0.95$.  }
\end{figure}

In Fig. \ref{fig:hom_no_ps}, we show the secret key rates as a function of the transmission distance for protocol 1 with homodyne detection for different choices of excess noise $\xi$. For this plot, we optimize the coherent state amplitude $\alpha$ by a coarse-grained search in the interval $[0.35,0.6]$. As we can see from the plot, we can reach around 200 km with an experimentally feasible value of excess noise, say, $\xi = 0.01$ \cite{Jouguet2013, Huang2016} with the current technology before the key rate becomes insignificant (say, less than $10^{-6}$ per pulse). To put the number in a more concrete and realistic context, if we consider a system with the repetition rate of 1 GHz and with the detector efficiency $55\%$, we can obtain $10^3$ bits per second at the distance of around 170 km if the total excess noise $\xi$ can be made to be 1\% or less.

\begin{figure}[h]
\subfloat[]{\label{fig:hom_ps_parameter}\includegraphics[width=0.48\linewidth, height=0.52\linewidth]{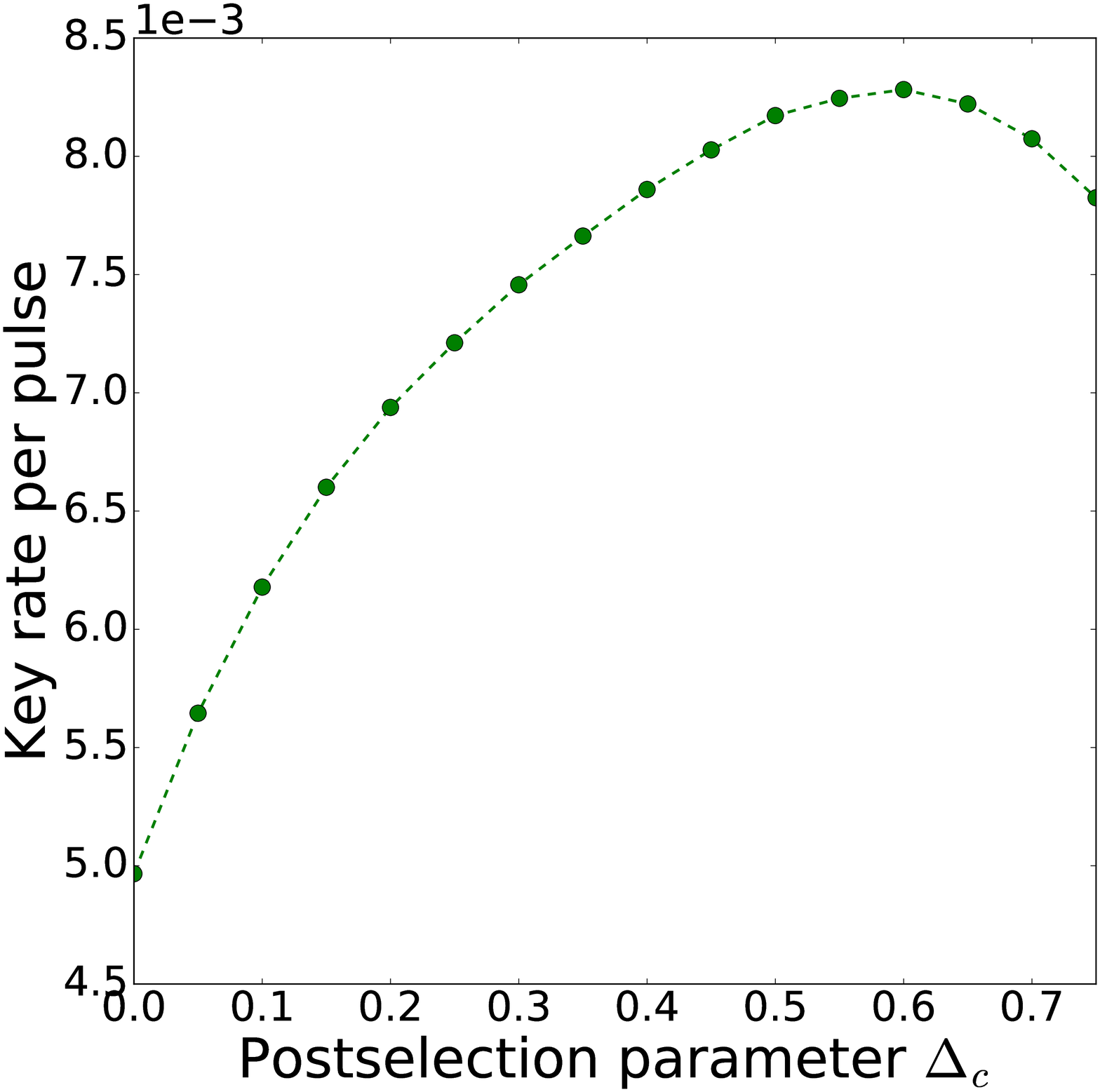}}
\subfloat[]{\label{fig:hom_ps_comp}\includegraphics[width=0.52\linewidth]{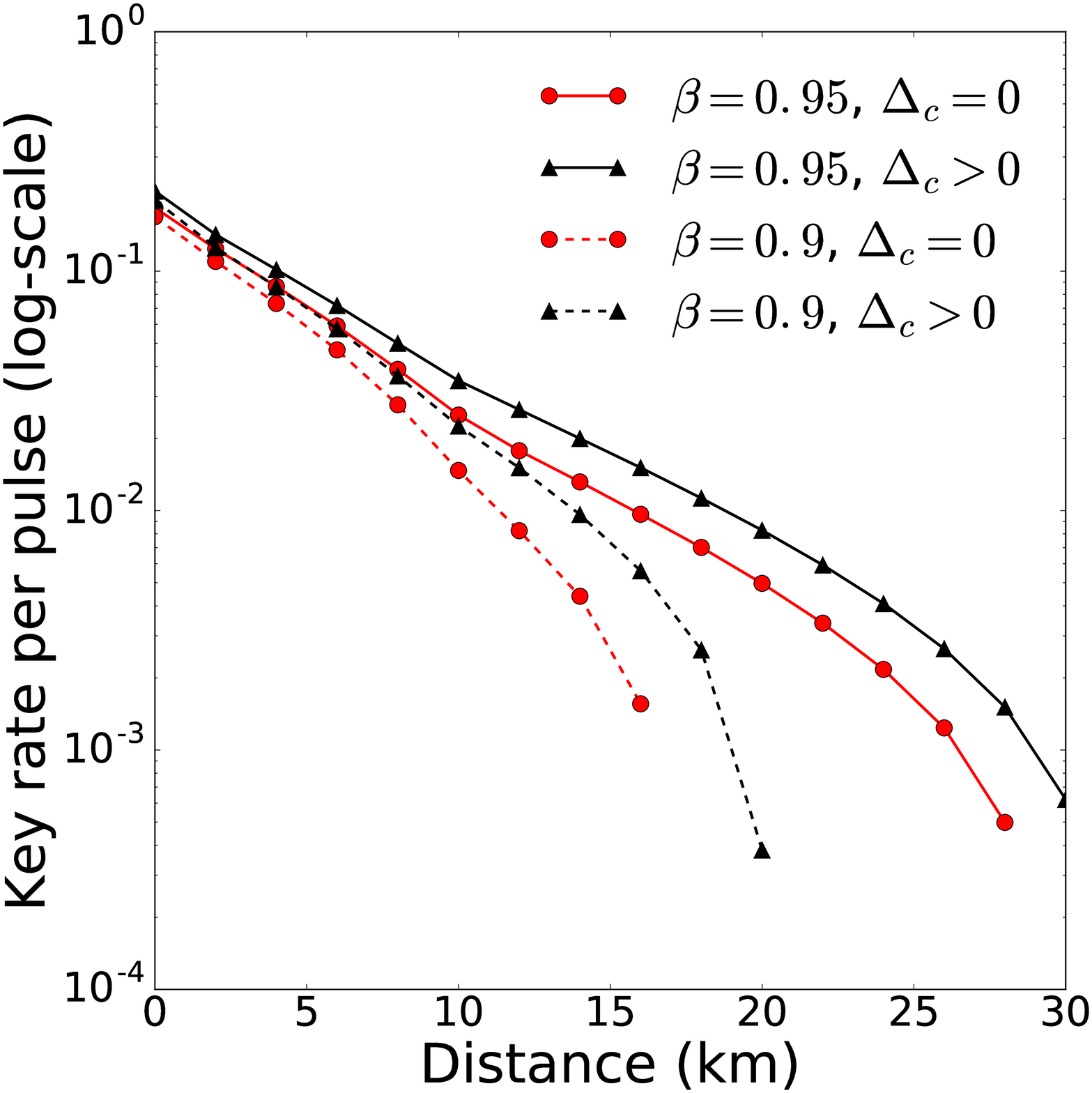}}
\caption{\label{fig:hom_ps} Secure key rate for protocol 1 (homodyne detection) with postselection. The excess noise is $\xi = 0.02$, and the coherent state amplitude is $\alpha= 0.45$. (a) Secure key rate versus the postselection parameter $\Delta_c$ at the distance $L=20$ km with the reconciliation efficiency $\beta= 0.95$. (b) Secure key rate versus the transmission distance with or without postselection for two different values of $\beta$. Solid lines have $\beta= 0.95$, and dashed lines have  $\beta= 0.9$. Lines with (red) circle markers have $\Delta_c = 0$, and lines with (black) triangle markers have $\Delta_c$ optimized via a coarse-grained search in the interval $[0.5,0.7]$. }
\end{figure}

We also investigate the effects of postselection. The idea of postselection was initially introduced to CV QKD protocols in order to beat the 3 dB limit \cite{Silberhorn2002}. The key rate can be potentially improved by discarding very noisy data where Eve has more advantages in determining the raw key than the party (Bob in the case of direct reconciliation and Alice in the case of reverse reconciliation) who needs to match the raw key via the error correction. Intuitively, if we optimize the postselection parameter $\Delta_c$, the key rate can never be lower than the protocol without postselection since one can always set $\Delta_c = 0$ if it is optimal to do so. The important observation here is that our security proof technique allows us to consider postselection with $\Delta_c >0$ by a simple modification of the postprocessing map $\mathcal{G}$, unlike previous security proofs based on Gaussian optimality. In Fig. \ref{fig:hom_ps}, we take the case with  an excess noise $\xi = 0.02$ and a fixed coherent state amplitude $\alpha= 0.45$ as an example to illustrate how the postselection strategy can improve the key rate in the reverse reconciliation scheme and to what extent it can help. We first search an optimal value for the postselection parameter $\Delta_c$ by a coarse-grained search, and we see, in Fig. \ref{fig:hom_ps_parameter}, that the optimal value is around 0.6 at the distance $L=20$ km. We also obtain similar plots for different choices of distance and find that the optimal value falls roughly in the interval $[0.5, 0.7]$. In Fig. \ref{fig:hom_ps_comp}, we compare the key rate with postselection ($\Delta_c > 0$) to that without postselection ($\Delta_c =0$) for two values of reconciliation efficiency $\beta$. In this plot, we optimize the postselection parameter $\Delta_c$ via a coarse-grained search in the interval $[0.5, 0.7]$. Since the curves with postselection are above the curves without postselection, we see that the postselection strategy can improve the key rates. We also notice that, for reverse reconciliation schemes, the advantage of postselection also depends on the reconciliation efficiency $\beta$. The gap between these two scenarios $\Delta_c = 0$ and $\Delta_c >0$ is smaller when a more efficient code (larger $\beta$) is used.

\subsection{Noisy scenario: Protocol 2}

\subsubsection{Simulated statistics and error-correction cost}
We now investigate protocol 2 which uses heterodyne detection. From the heterodyne measurements, for each conditional state $\rho_B^x$ with $\alpha_x \in \{\alpha ,i\alpha, -\alpha, -i\alpha\}$, we obtain a $Q$ function $Q_x$ as
\begin{aeq}
Q_x(\gamma) =  \frac{1}{\pi (1+\eta \xi / 2)} \exp(-\frac{\abs{\gamma-\sqrt{\eta}\alpha_x}^2}{1+\eta \xi / 2}).
\end{aeq}

From each $Q$ function, we can then calculate
\begin{aeq}
\langle \hat{q} \rangle_x  &= \frac{1}{\sqrt{2}}\int (\gamma + \gamma^*) Q_x(\gamma)  d^2 \gamma =  \sqrt{2\eta} \Re(\alpha_x),\\
\langle \hat{p} \rangle_x  &= \frac{i}{\sqrt{2}}\int (\gamma^* - \gamma) Q_x(\gamma)  d^2 \gamma =\sqrt{2\eta} \Im(\alpha_x),\\
\langle \hat{n} \rangle_x  &=\int ( \abs{\gamma}^2-1) Q_x(\gamma)  d^2 \gamma= \eta \abs{\alpha_x}^2 +\frac{\eta \xi}{2},\\
\langle \hat{d} \rangle_x  &=\int [\gamma^2 + (\gamma^*)^2] Q_x(\gamma)  d^2 \gamma = \eta [\alpha_x^2 +(\alpha_x^{*})^2].
\end{aeq}Note that those values are exactly the same as from the homodyne measurements, since we have the same state after the simulated quantum channel. We obtain those values here indirectly via the $Q$ function.

We also present the procedure to calculate $\delta_{\text{EC}}$ for protocol 2. We can numerically evaluate $H(\mathbf{Z}|\mathbf{X})$ via the probability distribution:
\begin{aeq}
& P(z=j | x=k) = \Tr(R_j \rho_B^k)\\
&= \int_{\Delta_a}^{\infty} \int_{\frac{2j-1}{4}\pi + \Delta_p}^{\frac{2j+1}{4}\pi -\Delta_p} \frac{\exp(-\frac{\abs{\gamma e^{i\theta }-\sqrt{\eta}\alpha_k}^2}{1+\eta \xi / 2})}{\pi (1+\eta \xi / 2)} \gamma \; d\theta \; d\gamma ,
\end{aeq}where $j,k \in \{0,1,2,3\}$, $R_j$'s are the region operators defined in Eq. (\ref{eq:region_operator}), and the conditional state $\rho_B^k$ is defined in Eq. (\ref{eq:conditional_state_B}). (In the case of postselection, we then renormalize this probability distribution by the probability of being postselected.) Then, $\delta_{\text{EC}}$ can be calculated by the second line of Eq. (\ref{eq:delta_ec}), as we now take into account that we have an alphabet of four symbols on both sides in the error-correction step.

\subsubsection{Key rates for protocol 2}

\begin{figure}[h]
\subfloat[$L= 20$ km]{\label{fig:het_alpha_d20}\includegraphics[width=0.5\linewidth]{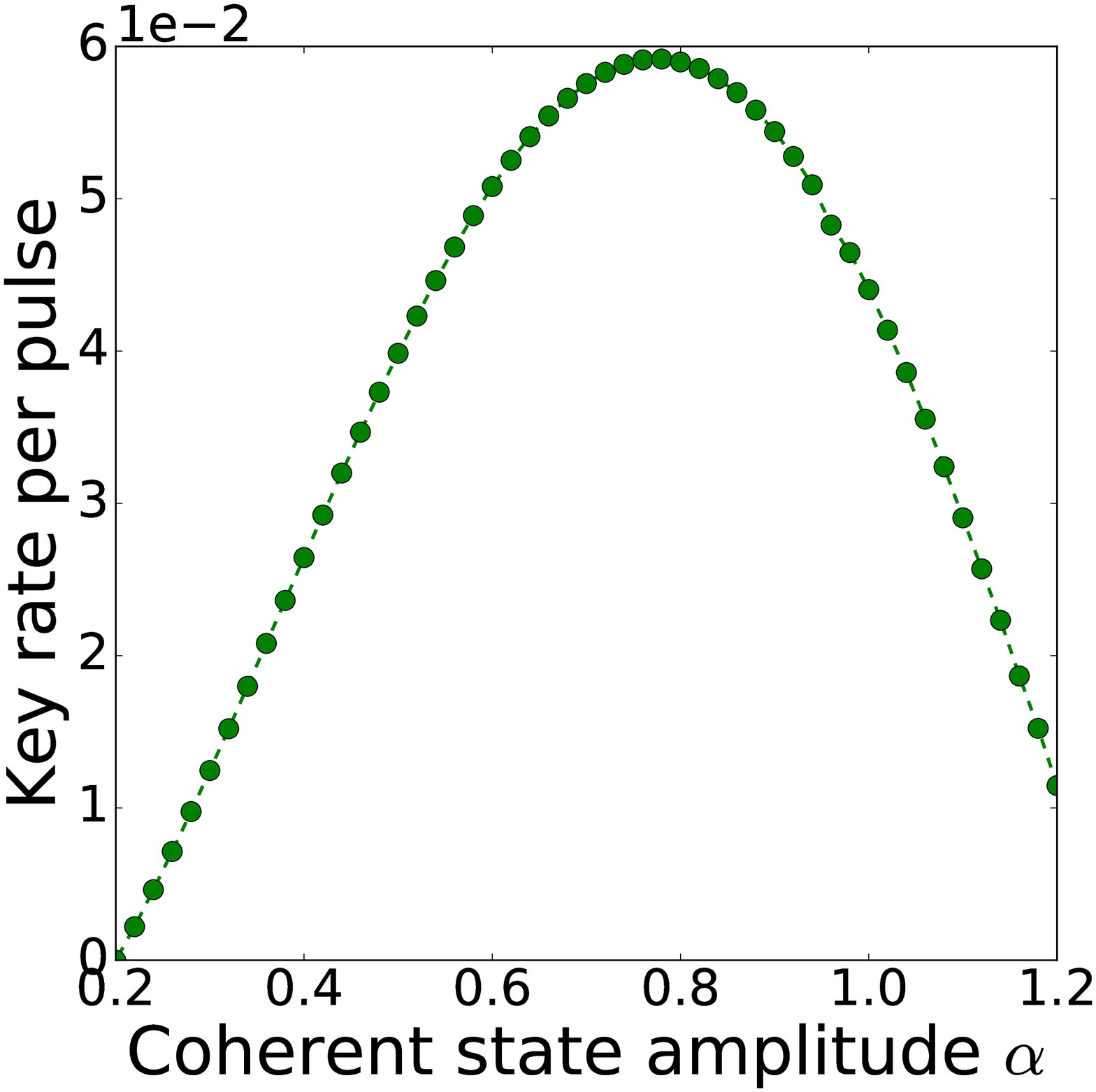}}
\subfloat[$L= 50$ km]{\label{fig:het_alpha_d50}\includegraphics[width=0.5\linewidth]{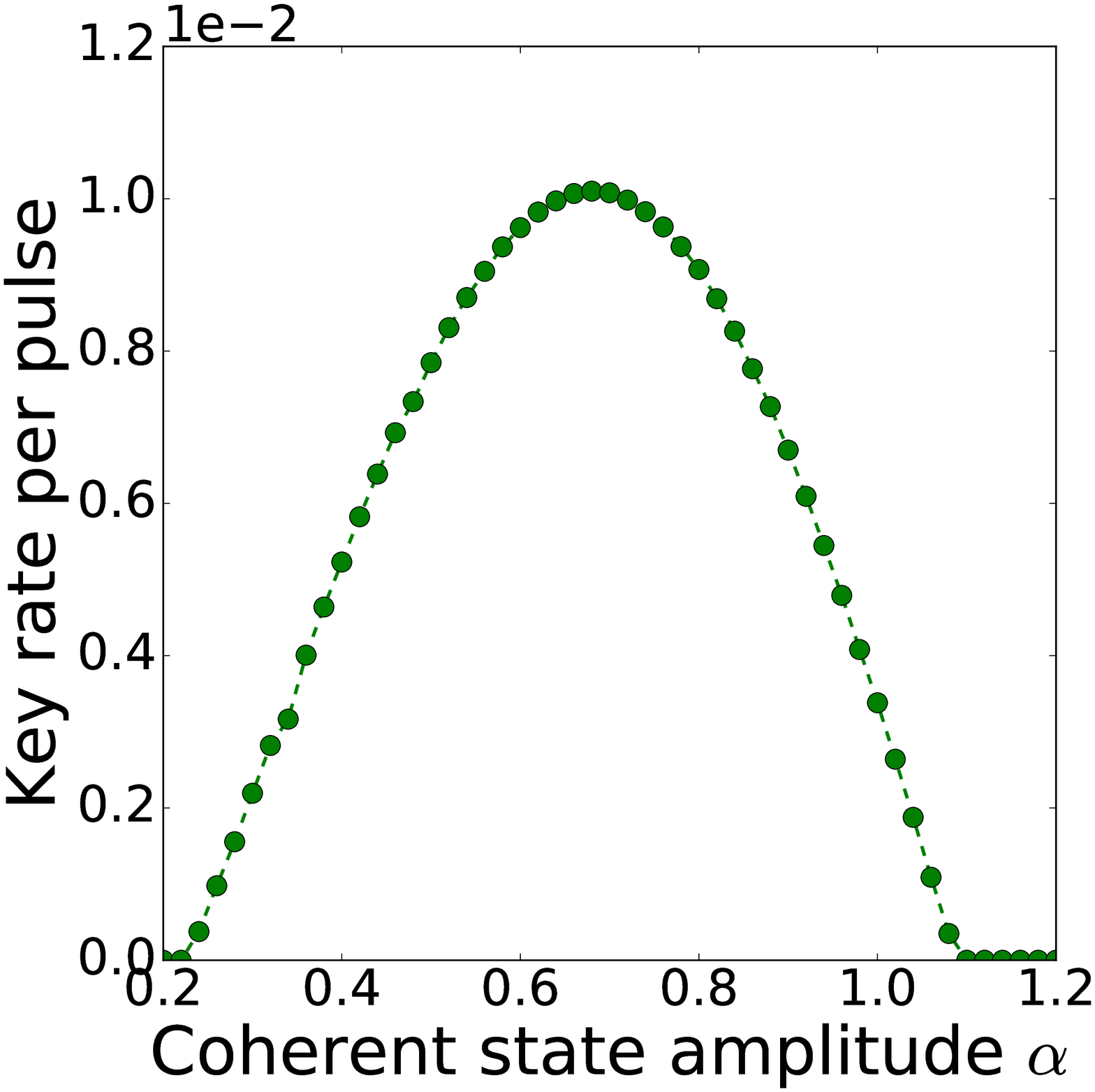}}\\
\subfloat[$L= 80$ km]{\label{fig:het_alpha_d80}\includegraphics[width=0.5\linewidth]{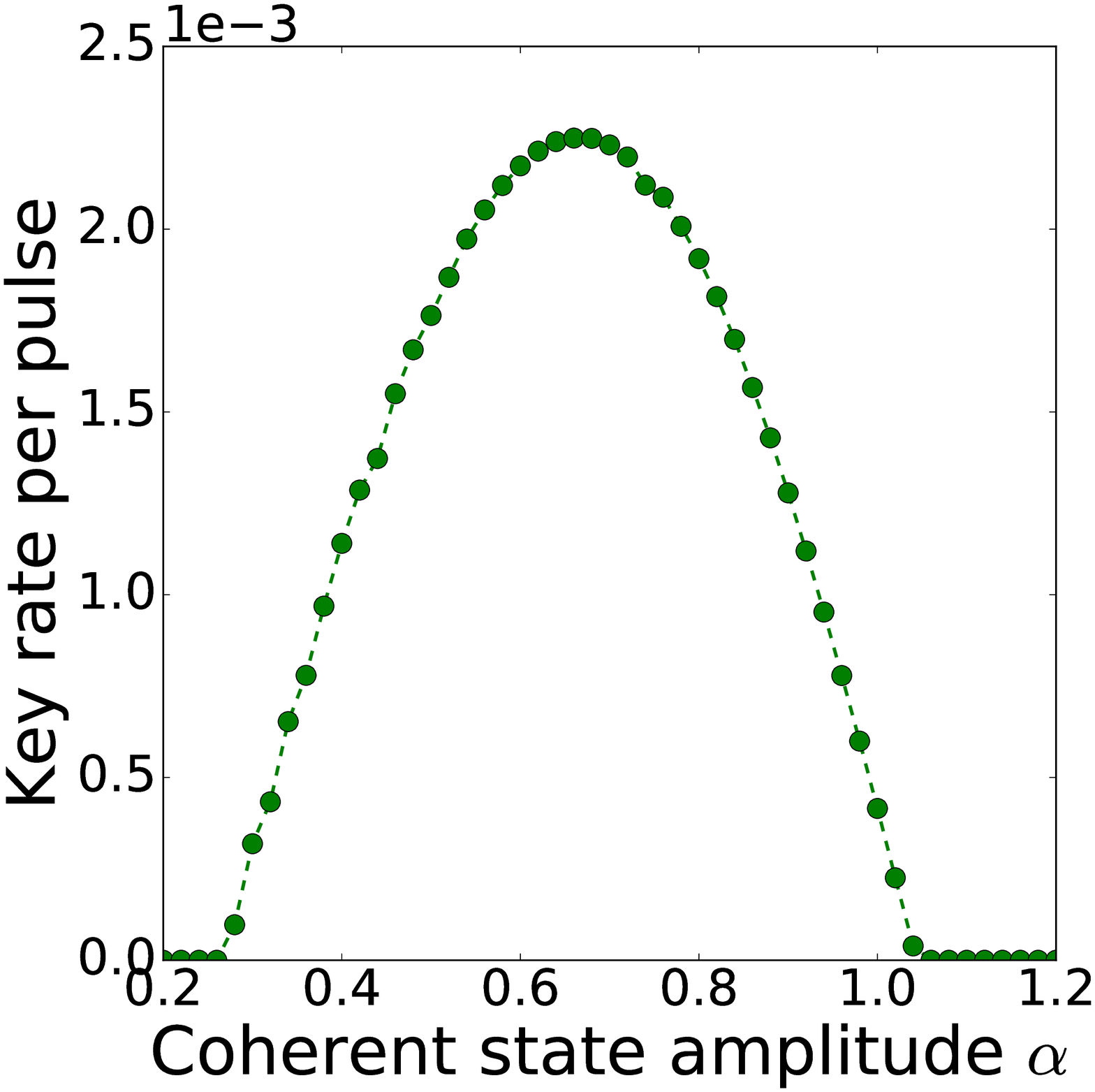}}
\subfloat[$L= 100$ km]{\label{fig:het_alpha_d100}\includegraphics[width=0.5\linewidth]{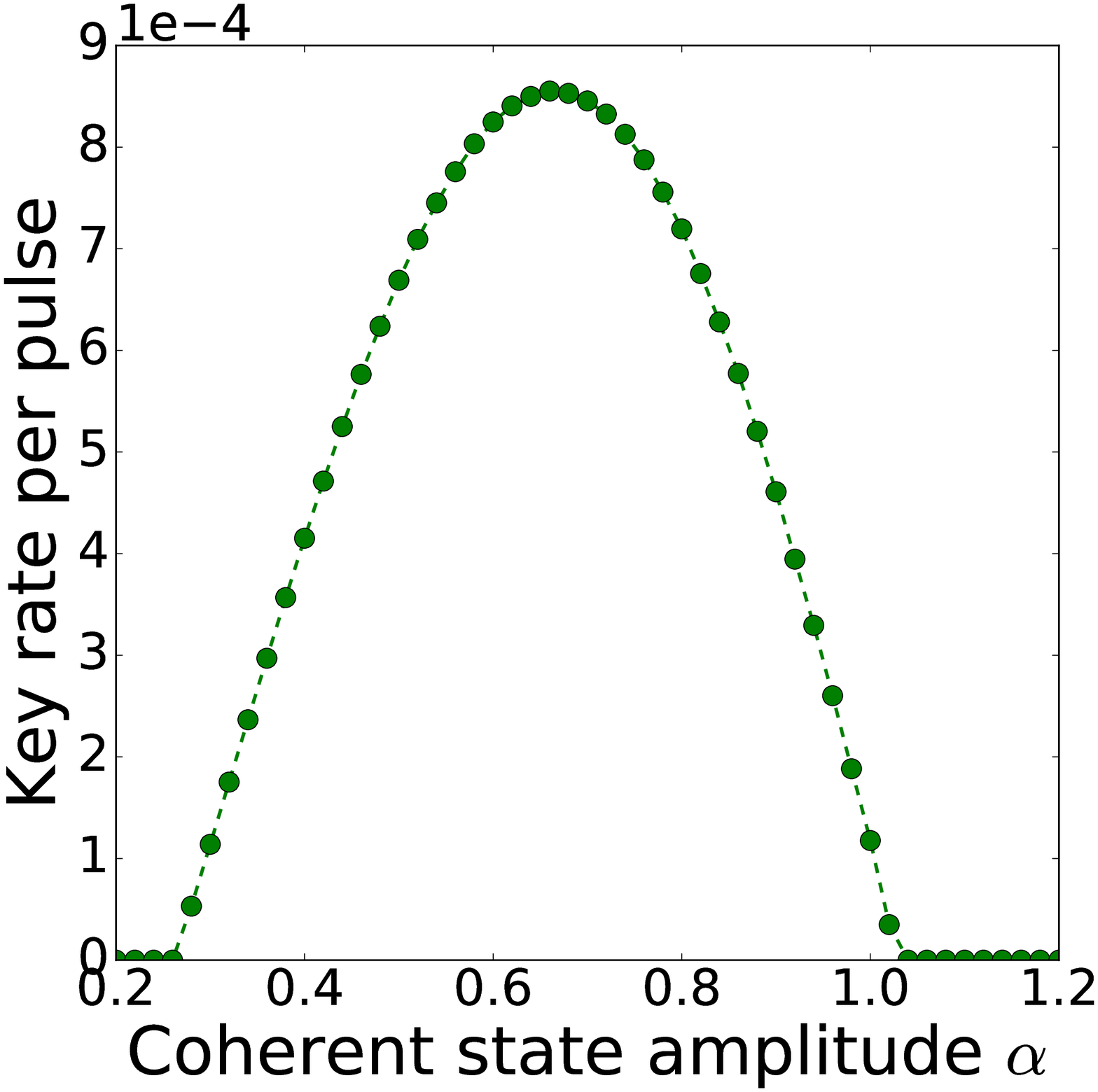}}
\caption{\label{fig:het_alpha} Secure key rate for protocol 2 versus the coherent state amplitude $\alpha$ for selected choices of distances  (a) $L = 20$ km, (b) $L = 50$ km, (c) $L = 80$ km, and (d) $L = 100$ km with the excess noise $\xi = 0.01$ and reconciliation efficiency $\beta = 0.95$. }
 \end{figure}
 
As with the case of protocol 1, we start by investigating the optimal choice of coherent state amplitude $\alpha$ for protocol 2. In Fig. \ref{fig:het_alpha}, we plot the key rates versus $\alpha$ for selected distances when the excess noise $\xi$ is 0.01. Comparing to Fig. \ref{fig:hom_alpha}, we see that the optimal $\alpha$ for this variant with heterodyne detection is, in general, larger than that for protocol 1. The optimal choice of $\alpha$ in protocol 2 is around 0.7 for those selected distances, corresponding to a mean photon number around 0.49, while the optimal choice of $\alpha$ in protocol 1 is around 0.4 for those selected distances, corresponding to a mean photon number around 0.16. Like protocol 1, we observe that the optimal value of $\alpha$ for protocol 2 does not change significantly for a wide range of distances. From the observation here, we later limit our search for optimal choice of $\alpha$ in a restricted interval.

\begin{figure}[h]
\includegraphics[width=\linewidth]{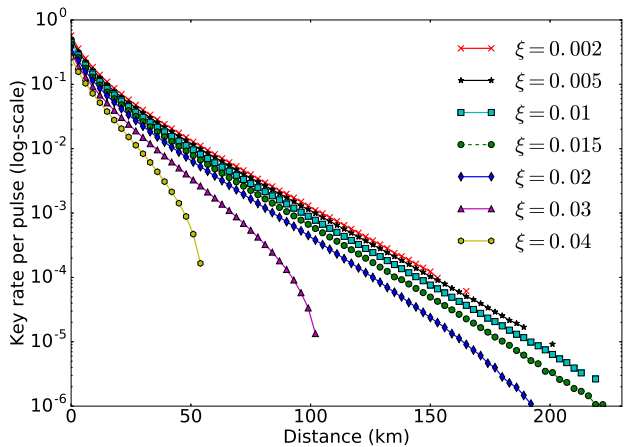}
\caption{\label{fig:het_no_ps} Secure key rate versus the transmission distance for protocol 2 with heterodyne detection for different values of the excess noise $\xi$, from top to bottom, $\xi = 0.002, 0.005, 0.01, 0.015, 0.02, 0.03, 0.04$. The coherent state amplitude is optimized via a coarse-grained search over the interval $[0.6,0.92]$, the transmittance is $\eta = 10^{-0.02 L}$ for each distance $L$ in kilometers and the reconciliation efficiency is $\beta = 0.95$. }
\end{figure}

\begin{figure}[h]
\includegraphics[width=\linewidth]{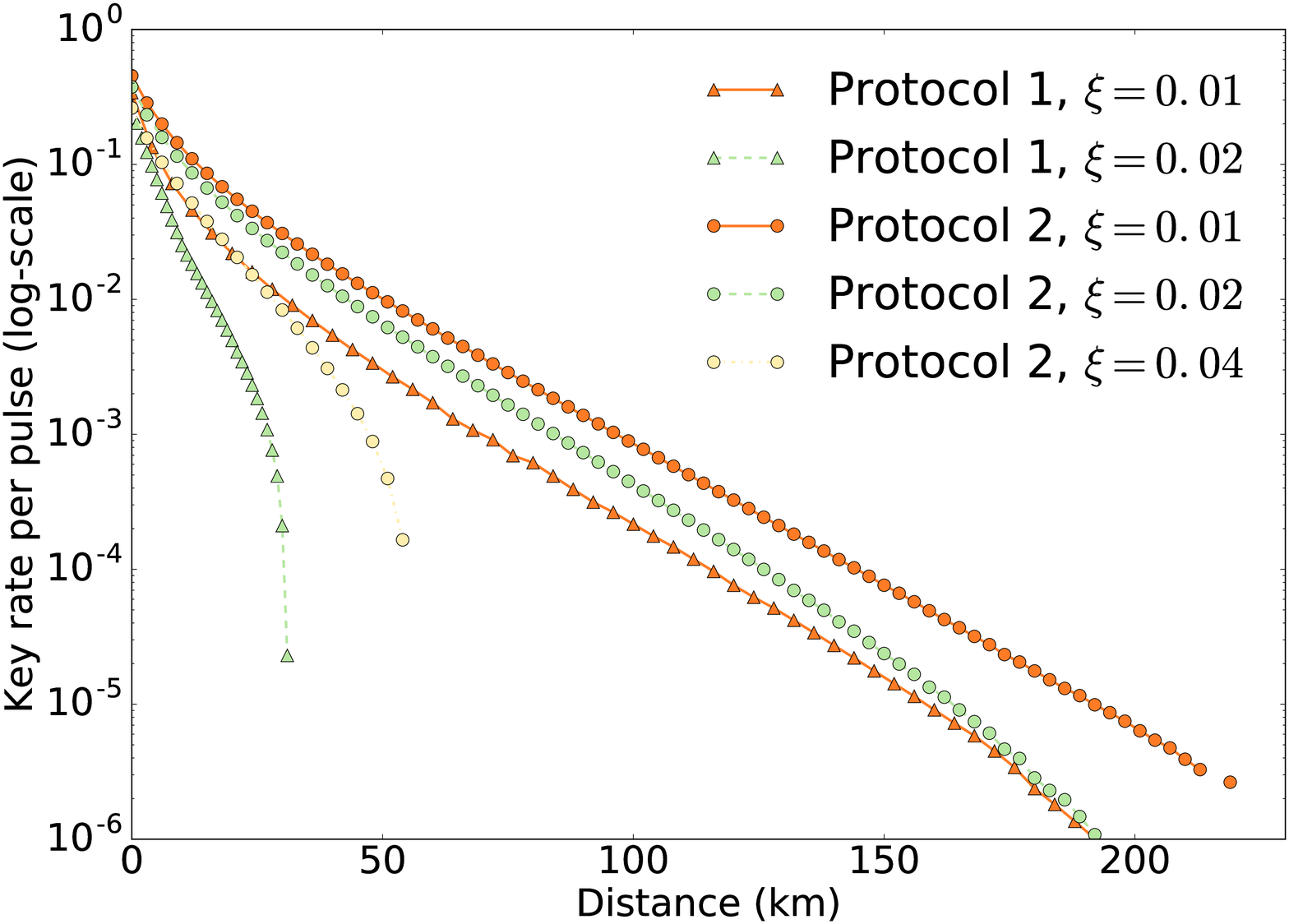}
\caption{\label{fig:comparison} Secure key rate versus transmission distance for a direct comparison between protocols 1 and 2. Curves for protocol 1 are plotted with triangle markers (from Fig. \ref{fig:hom_no_ps}); the excess noise is $\xi = 0.01, 0.02$ from top to bottom for curves with triangle markers. Curves for protocol 2 are plotted with circle markers (from Fig. \ref{fig:het_no_ps}); the excess noise is $\xi = 0.01, 0.02, 0.04$ from top to bottom for curves with circle markers.}
\end{figure}

In Fig. \ref{fig:het_no_ps}, we plot the secure key rate versus the transmission distance for different values of excess noise $\xi$. We optimize the coherent state amplitude $\alpha$ via a coarse-grained search over the interval $[0.6,0.92]$. We observe that protocol 2 can reach around 200 km even with an excess noise $\xi = 0.02$. Interestingly, we see that the key rate for protocol 2 is much higher than protocol 1 when the excess noise is large. For a direct comparison, we replot key rates of both protocols for the values of excess noise $\xi = 0.01$ and $0.02$ from Figs. \ref{fig:hom_no_ps} and \ref{fig:het_no_ps} in Fig. \ref{fig:comparison}. We observe that protocol 2 achieves much higher key rates and reaches longer distances than protocol 1 for the same amount of excess noise. In this figure, we also plot the key rate of protocol 2 with the excess noise $\xi = 0.04$ for a direct comparison to the key rate of protocol 1 with the excess noise $\xi = 0.02$. We see that protocol 2 behaves similarly as protocol 1 with half of the excess noise for those values of excess noise considered here.

\begin{figure}[h]
\subfloat[$\xi= 0.005$]{\label{fig:hetcomp_005}\includegraphics[width=\linewidth]{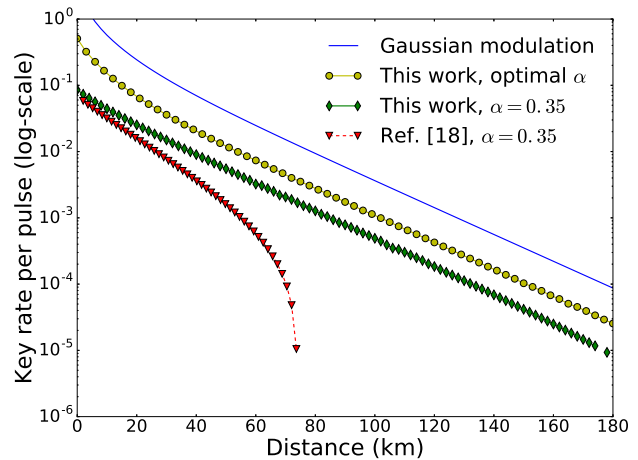}}\\
\subfloat[$\xi= 0.01$]{\label{fig:hetcomp_010}\includegraphics[width=\linewidth]{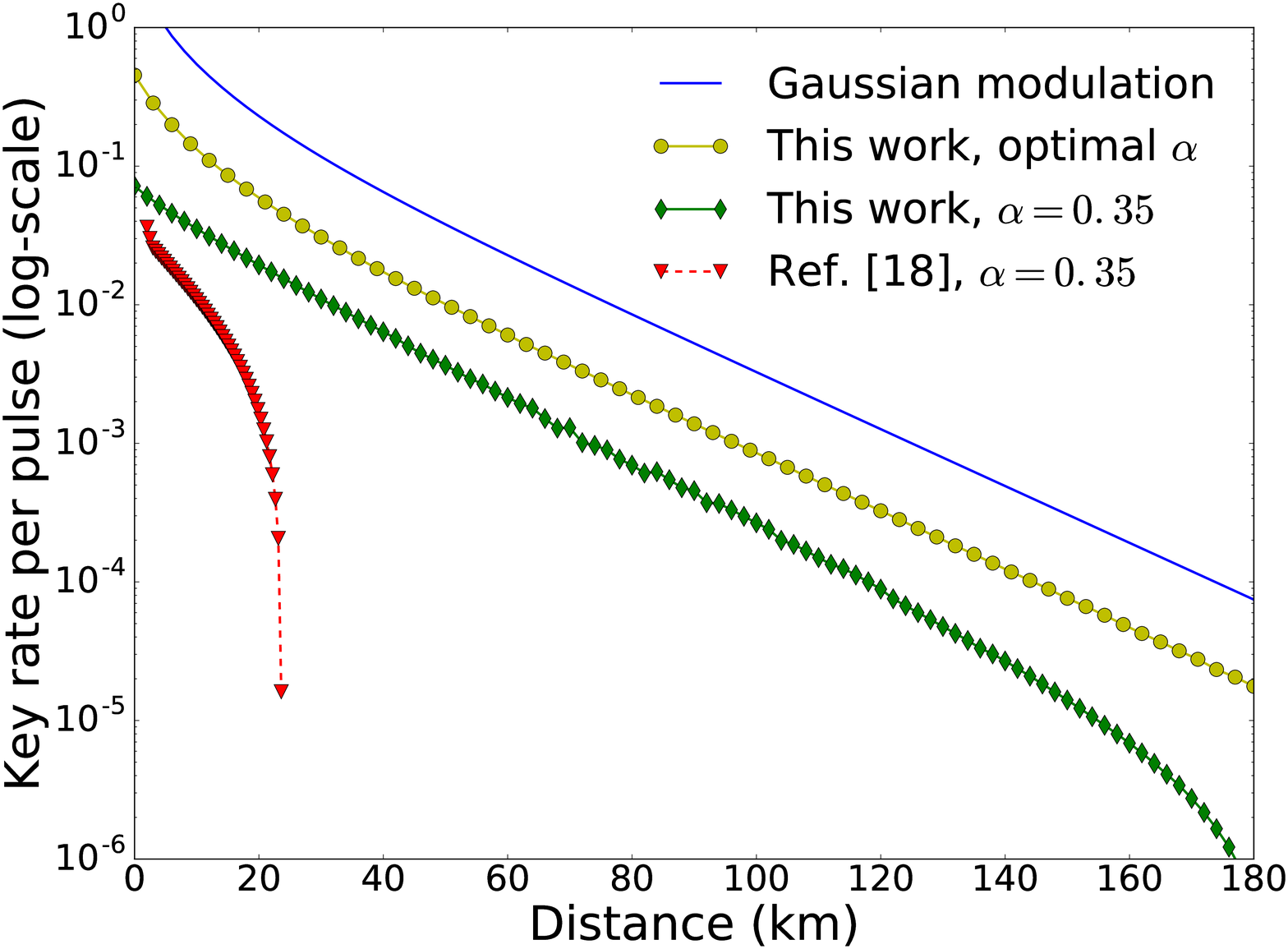}}
\caption{\label{fig:different_work_comparison} A comparison of key rates between our work and Ref. \cite{Ghorai2019} for the heterodyne scheme with two different values of excess noise: (a) $\xi = 0.005$ and (b) $\xi = 0.01$. In each plot, the curve with triangle markers is from Ref. \cite{Ghorai2019} with a fixed (not optimal) coherent state amplitude $\alpha = 0.35$, and the curve with diamond markers is from this work with the same value of $\alpha$. The curve with circle markers is also from this work with an optimal value of $\alpha$ for each distance. The curve with no markers is the key rates of a Gaussian-modulated CV QKD protocol \cite{Weedbrook2004} with an optimal modulation variance for each distance.  All curves use the reconciliation efficiency $\beta = 0.95$. }
\end{figure}

We then compare our results with an independent security analysis in Ref. \cite{Ghorai2019} for a similar protocol. In addition to different proof methods, we differ from that protocol by how the error correction is done, which affects the calculation of the error-correction cost term $\delta_{\text{EC}}$. In particular, our error-correction cost is higher because we discretize Bob's measurement results and consider only binary or quaternary error-correcting codes. In Ref. \cite{Ghorai2019}, the mutual information $I(\mathbf{X};\mathbf{Z})$ is obtained by the channel capacity of the binary additive white Gaussian noise channel, which is approximated by the capacity of an additive white Gaussian noise channel
\begin{equation}
I(\mathbf{X}; \mathbf{Z}) \approx \log_2 \Big( 1+ \frac{2 \eta \alpha^2}{ 2+ \eta \xi}\Big).
\end{equation} This result leads to a smaller value of $\delta_{\text{EC}}$ by the conversion formula in the first line of Eq. (\ref{eq:delta_ec}). In Fig. \ref{fig:different_work_comparison}, we plot the key rate results from both our work and Ref. \cite{Ghorai2019} with a fixed choice of the coherent state amplitude $\alpha =0.35$ for all distances plotted and with two different values of excess noise. As we can see, our security proof approach provides remarkably higher key rates compared with the approach with a reduction to the Gaussian optimality. Our security analysis shows that this protocol has a good tolerance on the excess noise and can extend to significantly longer distances. We emphasize that this choice of $\alpha = 0.35$ is not optimal for both works. While this value is closer to the optimal value found in Ref. \cite{Ghorai2019}, the optimal value of the coherent state amplitude found in our work is around 0.7 (for $\xi = 0.01$), as mentioned before. Thus, we also include two curves from Fig. \ref{fig:het_no_ps}, where the coherent state amplitude $\alpha$ is optimized via a coarse-grained search in the interval $[0.6,0.92]$ for comparisons. As we can see from Fig. \ref{fig:different_work_comparison}, the key rate can be significantly improved after we optimize $\alpha$. We summarize two factors that can boost the key rates. First, our security proof technique gives a tighter estimation of Eve's information compared with the reduction to the Gaussian optimality approach. Second, the key rate can be improved by using a slightly larger value of $\alpha$ than what is investigated in Ref. \cite{Ghorai2019}. This regime of $\alpha$ is not explored in Ref. \cite{Ghorai2019}, because the reduction to the Gaussian optimality approach for discrete-modulation schemes gives tight key rates only in the limit of $\alpha \rightarrow 0$ and can give quite loose key rates for large values of $\alpha$. In the same figure, we also compare our results for protocol 2 with a Gaussian-modulated CV QKD protocol using heterodyne detection \cite{Weedbrook2004}, where the modulation variance is optimized for each distance. We observe that this quaternary modulation scheme has key rates comparable to the Gaussian modulation scheme.

\begin{figure}[h]
\subfloat[]{\label{fig:het_ps_parameter}\includegraphics[width=0.48\linewidth, height=0.52\linewidth]{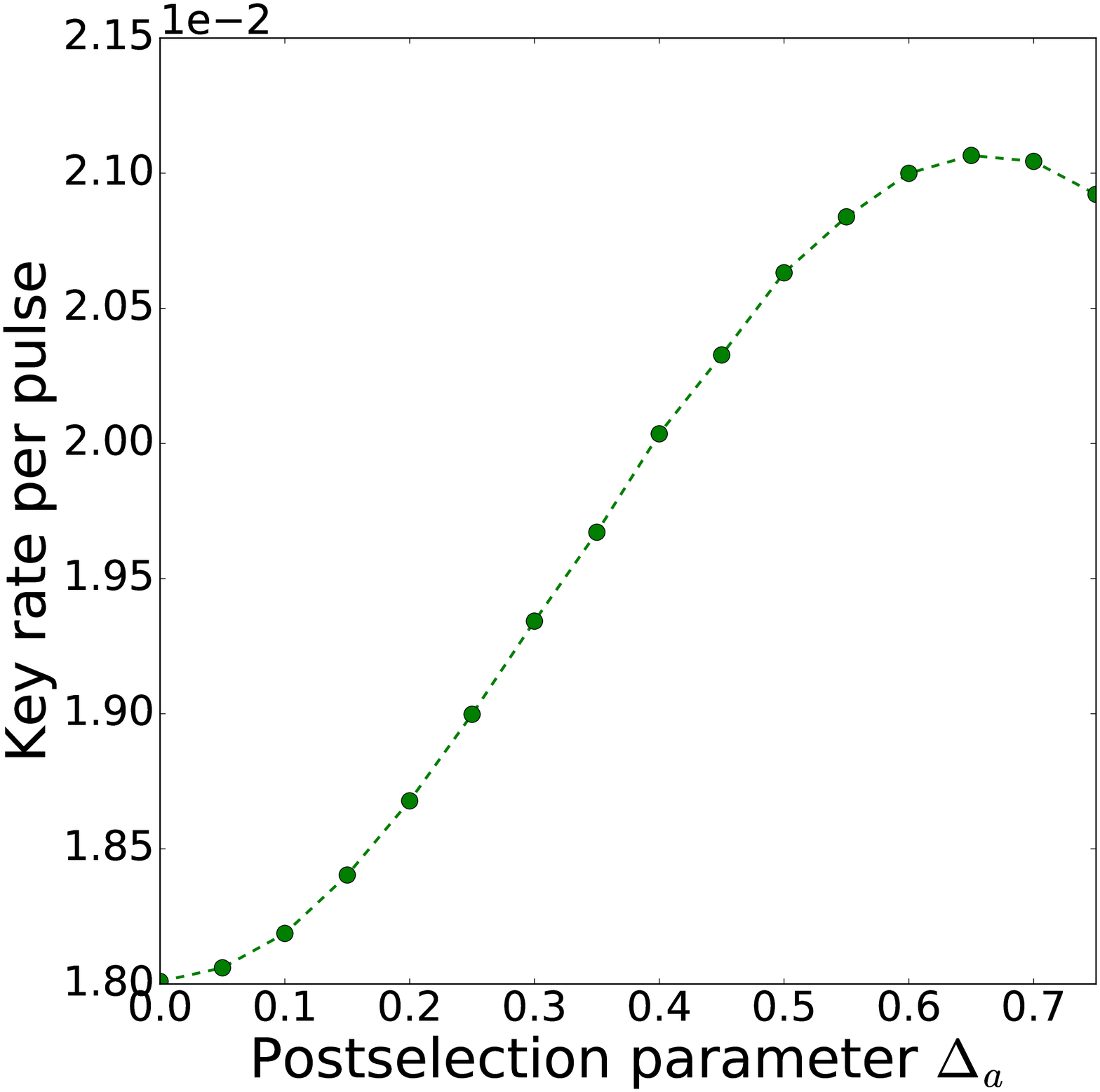}}
\subfloat[]{\label{fig:het_ps_comp}\includegraphics[width=0.52\linewidth]{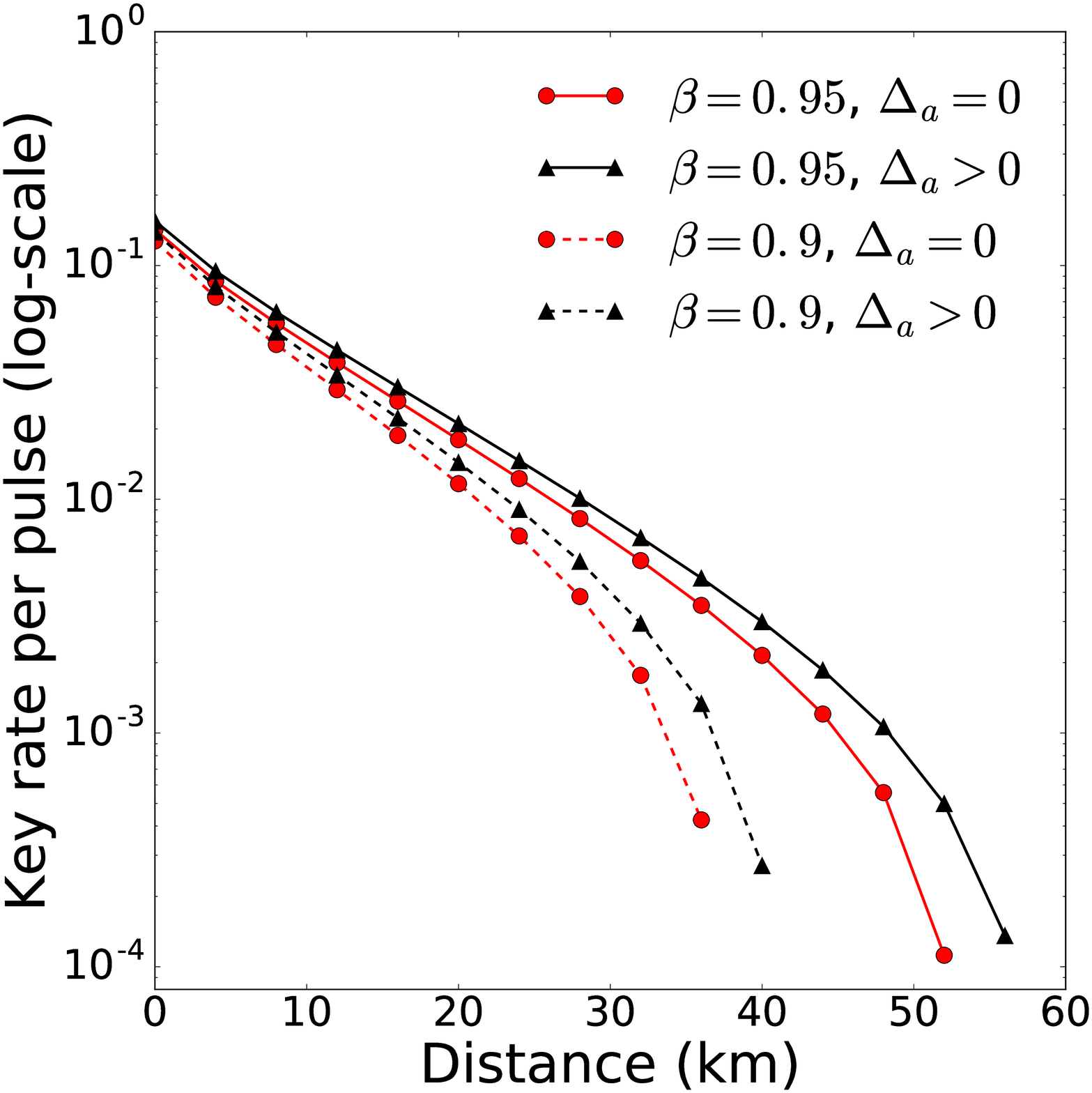}}
\caption{\label{fig:het_ps} Secure key rate for protocol 2 (heterodyne detection) with postselection. The excess noise is $\xi = 0.04$, the coherent state amplitude is $\alpha= 0.6$, and one of the postselection parameters is $\Delta_p=0$. (a) Secure key rate versus the postselection parameter $\Delta_a$ at the distance $L=20$ km. The reconciliation efficiency is $\beta= 0.95$. (b) Secure key rate versus the transmission distance with or without postselection for two different values of reconciliation efficiency $\beta$.  Solid lines have $\beta= 0.95$ and dashed lines have $\beta= 0.9$. Lines with (red) circle markers have $\Delta_a = 0$, and lines with (black) triangle markers have $\Delta_a$ optimized via a coarse-grained search in the interval $[0.4,0.7]$. }
\end{figure}

Finally, we present the results on the effects of postselection. Our coarse-grained search for values of $\Delta_p$ suggests that the optimal value is $\Delta_p=0$; that is, we do not postselect the data based on the phase. For the postselection parameter $\Delta_a$ related to the amplitude of the measured complex value from heterodyne detection, we then perform a coarse-grained search for its optimal value. In Fig. \ref{fig:het_ps}, we consider the scenario with an excess noise $\xi=0.04$ and with a fixed coherent state amplitude $\alpha=0.6$ as an example. In Fig. \ref{fig:het_ps_parameter}, we plot the key rate versus this parameter $\Delta_a$ at the distance $L=20$ km with the reconciliation efficiency $\beta = 0.95$. From this plot, we observe that the optimal value of $\Delta_a$ is around 0.6 at this distance. We also obtain similar plots for various choices of the distance and find that the optimal value roughly falls in the interval $[0.4,0.7]$. In Fig. \ref{fig:het_ps_comp}, we compare key rates with or without postselection for two different values of reconciliation efficiency at different transmission distances, and, for this plot, we optimize the values of $\Delta_a$ via a coarse-grained search in the interval $[0.4,0.7]$. We again notice that postselection with reverse reconciliation can improve the key rates. We remark that the improvement due to postselection in the reverse reconciliation scheme is more visible for less efficient error-correcting codes, larger excess noise, and longer transmission distances. This result agrees with the observation made in Ref. \cite{Heid2006} under a restricted class of attacks. 

\section{Summary and Outlook}\label{sec:outlook}

In this work, we investigate the asymptotic security of discrete-modulated CV QKD protocols against collective attacks and demonstrate our proof method on the quadrature phase-shift keying scheme.  We observe that CV QKD with quaternary modulation can significantly improve the key rate compared with previous binary and ternary modulation schemes \cite{Zhao2009, Bradler2018}. We also directly compare our results with the results in a recent independent work \cite{Ghorai2019}, which uses a different proof technique from ours. Interestingly, as our security proof approach can give tight key rates, we are able to obtain significantly higher key rates than Ref. \cite{Ghorai2019}. Our results show that this protocol can achieve comparable key rates as Gaussian modulation schemes. In addition, we consider the effects of postselection and demonstrate that postselection can improve the key rates. 

We remark on possible directions for future work. Since our security analysis imposes a photon-number cutoff assumption which truncates the total dimension of the system by ignoring the subspace that has negligible contributions, it requires further investigations to remove this assumption and to generalize the current proof to include finite-size effects and general attacks. We mention two possible paths to reach this final goal in Sec. \ref{sec:cutoff_assumption}, and there are challenges in each direction. In addition, there are also possible ways to improve the key rates. In both protocols, we discretize Bob's measurement outcomes to obtain binary or quaternary strings. Therefore, we consider only binary or quaternary error-correcting codes. One can potentially improve the key rates by a better choice of key map and, thus, a better error-correction strategy. For a better choice of key map, one may run the optimization in Eq. (\ref{eq:optimization}) with modified $\mathcal{G}$ and $\mathcal{Z}$ maps. Another direction for improvement is to find a better way to treat the imperfection of detectors in the security analysis. We currently treat detectors as perfect detectors. To obtain key rates with imperfect detectors, we can apply a simple but pessimistic treatment; that is, additional loss and additional excess noise due to imperfect detectors are attributed to Eve. By doing so, we can simply modify two parameters $\eta$ and $\xi$ to obtain key rates related to imperfect detectors. However, since detectors are securely located in Bob's laboratory, one may treat the excess noise due to detector imperfection like the electronic noise as trusted noises (e.g., see Refs. \cite{Lodewyck2007,Jouguet2013, Huang2016}). By not giving Eve this additional power, one may improve the key rates. We leave further improvements to future work.

\begin{acknowledgments}
We thank Patrick Coles and Adam Winick for valuable discussions on the numerical method in Ref. \cite{Winick2018} and their contributions to the MATLAB codes related to Ref. \cite{Winick2018} and thank Ian George for code review. We thank Anthony Leverrier for helpful discussions and comments on an early version of this manuscript and for providing data points from Ref. \cite{Ghorai2019} for the comparison in Fig. \ref{fig:different_work_comparison}.  We thank Christoph Marquardt and Kevin Jaksch for discussion about feasible experimental parameters. The work has been performed at the Institute for Quantum Computing, University of Waterloo, which is supported by Industry Canada. The research has been supported by NSERC under the Discovery Grants Program, Grant No. 341495, and under the Collaborative Research and Development Program, Grant No. CRDP J 522308-17. Financial support for this work has been partially provided by Huawei Technologies Canada Co., Ltd.
\end{acknowledgments}

\appendix

\section{FRAMEWORK FOR POSTPROCESSING: DERIVATION AND SIMPLIFICATION}\label{app:postprocessing}

\begin{figure}
\subfloat[]{\includegraphics[width=0.9\linewidth]{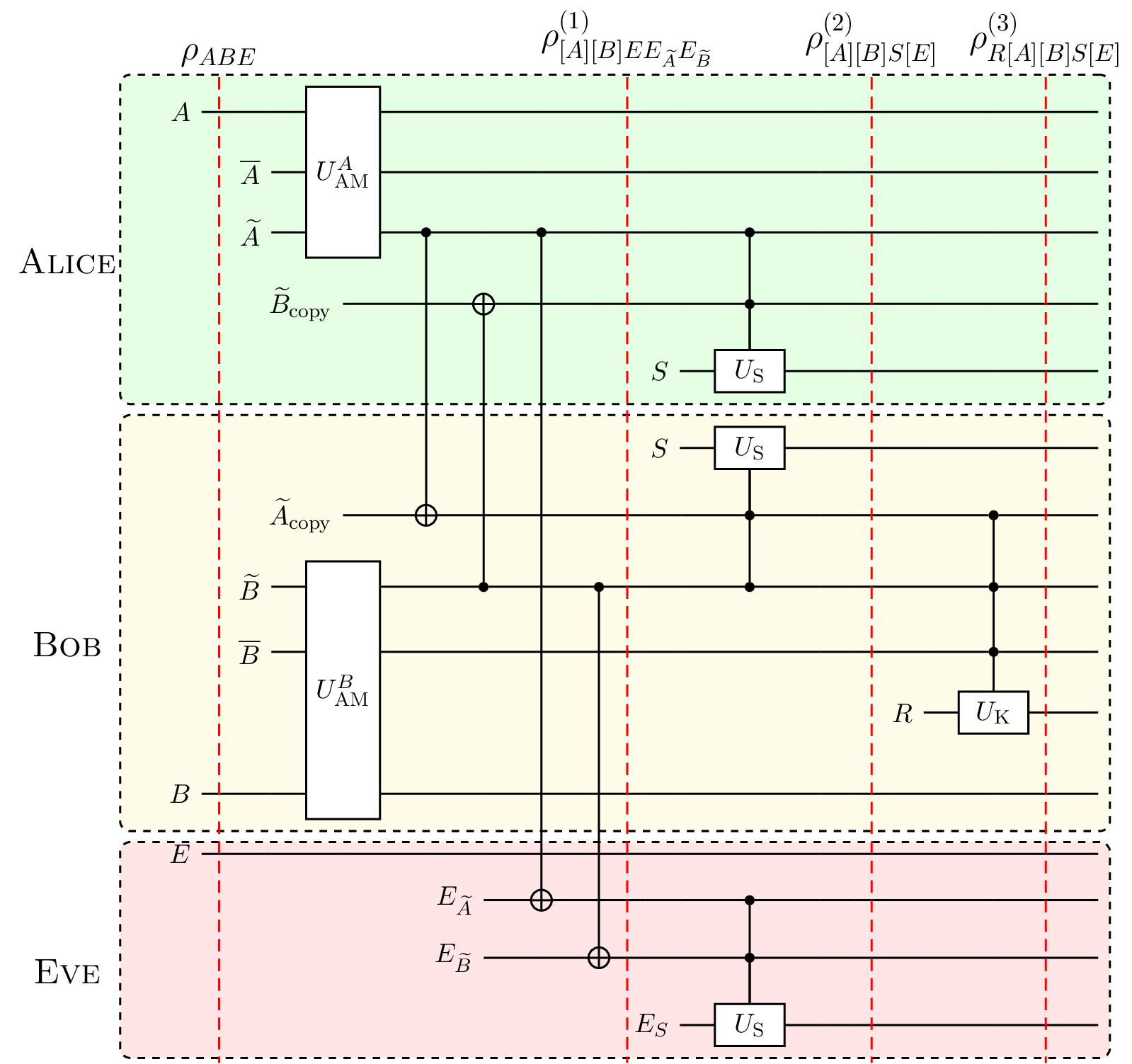}}\\
\subfloat[]{\includegraphics[width=0.8\linewidth]{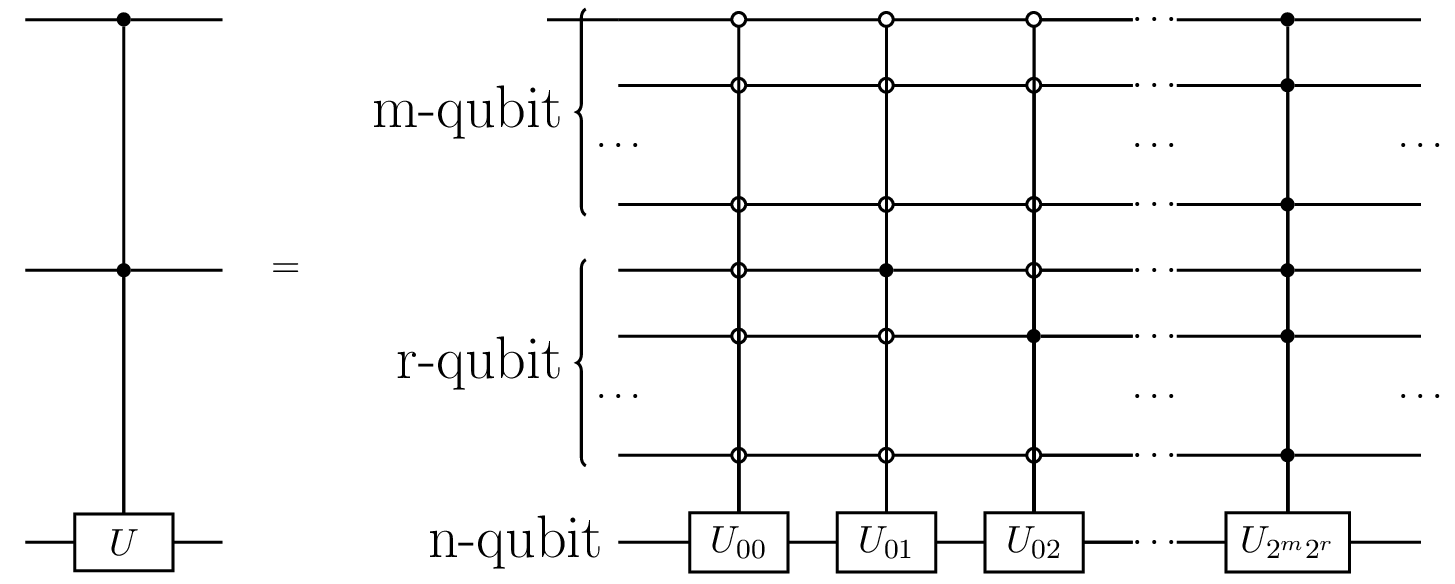}}
\caption{\label{fig:pp_full}  (a) A schematic circuit diagram for the relevant postprocessing steps: announcement (with measurement), sifting, and key map. Three dashed boxes separate Alice's, Bob's, and Eve's domains. The initial pure state $\rho_{ABE}$ is evolved by an isometry at each step which introduces additional registers and applies a unitary operation on relevant registers. $U^A_{\text{AM}}, U^B_{\text{AM}}, U_{\text{S}}$ and $U_{\text{K}}$ are unitary operations related to the announcement and measurement, sifting, and key map. (b) An explanation for the controlled unitary operation used in (a). Here, $m, r$, and $n$ are some sufficiently large integers so that we have a representation of the basis elements for each register in the computational basis of qubits. See the text in Appendix \ref{app:postprocessing} for more explanations. }
\end{figure}

In this Appendix, we present a derivation of a general postprocessing framework for the numerical method in Ref. \cite{Winick2018}, which reproduces the form of postprocessing map $\mathcal{G}$ presented in Ref. \cite{Winick2018} when Eve's systems are traced out. Based on this derivation, we then make several observations to simplify the postprocessing map $\mathcal{G}$ in special cases, which leads to a reduction of the dimensions required in the numerical analysis. Finally, we remark how $\mathcal{G}$ maps are simplified in this work.

We begin with some definitions to set up the notations. When we write an operator on composite registers, we omit the identity operator on the unspecified registers and may reorder registers for ease of writing. Moreover, the relevant unspecified registers depend on the context. Let $\mathcal{X}$ denote the set of Alice's measurement outcomes. With Alice's set of announcements $S^A$, we partition the set $\mathcal{X}$ into subsets $\mathcal{X}_a$ for $a \in S^A$ such that $\mathcal{X} = \bigcup_{a \in S^A} \mathcal{X}_a$.  Similarly, we partition Bob's measurement outcomes $\mathcal{Y}$ as $\mathcal{Y} =\bigcup_{b \in S^B} \mathcal{Y}_b$ by using his set of announcements $S^B$. To simplify our notation, we assume without the loss of generality that $\abs{\mathcal{X}_a} = \omega_A$ for all $a \in S^A$ and $\abs{\mathcal{Y}_b} = \omega_B$ for all $b \in S^B$ for some numbers $\omega_A$ and $\omega_B$ independent of $a$ and $b$. This assumption can be easily satisfied by a clever way of bookkeeping measurement outcomes. Then, we define a family of maps $f_a: \Omega^A:=\{1,2,\dots, \omega_A\} \rightarrow \mathcal{X}_a$ and a family of maps $f_b: \Omega^B:= \{1,2,\dots, \omega_B\} \rightarrow  \mathcal{Y}_b $ such that $f_a$'s and $f_b$'s are bijective. Finally, we label Alice's POVM $P^A$ as $P^A:= \{P^A_x: x \in \mathcal{X}\} = \{P^A_{[a, f_a(\alpha)]}: a \in S^A, \alpha \in \Omega^A\}$ and Bob's POVM $P^B := \{P^B_y: y \in \mathcal{Y}\} = \{P^B_{[b, f_b(\beta)]}: b \in S^B, \beta \in \Omega^B\}$.

\subsection{A full model for the relevant postprocessing steps}

We give a schematic circuit diagram in Fig. \ref{fig:pp_full} to describe the announcement (including measurement), sifting and key map steps. This diagram covers the scenarios related to this work, and it works for protocols with one round of announcements and with reverse reconciliation. It is not difficult to draw a similar diagram in other scenarios, including the direct reconciliation schemes. Under collective attacks, Alice and Bob share a bipartite quantum state $\rho_{AB}$ after each transmission of a quantum signal. In the worst-case scenario, Eve holds a purification $\rho_{ABE}$ of $\rho_{AB}$.  The initial state in the circuit diagram is $\rho_{ABE}$. At each step, the state is evolved by an isometry; that is, we introduce some local registers and evolve the state by a local unitary. We also keep track of the information leakage during the classical communication.  If some information is publicly available during the classical communication in the protocol, then each party holds a copy of the relevant registers. One can recover the classical communication information by measuring a local copy of these registers in the computational basis. Now, we discuss each of the three steps in detail. 

The isometries related to the announcement and measurement step are denoted by $W_A$ for Alice and  $W_B$ for Bob. In particular, $W_A$ first introduces two registers $\widetilde{A}$ and $\overline{A}$ and then applies a unitary operator $U^A_{\text{AM}}$ to implement Alice's POVM $P^A$ in a coherent fashion, where announcements are stored in the register $\widetilde{A}$ and measurement outcomes are in the register $\overline{A}$. Like Alice's isometry $W_A$, Bob's isometry $W_B$ implements his POVM $P^B$ with the announcement register $\widetilde{B}$ and measurement outcome register $\overline{B}$ using a local unitary $U^B_{\text{AM}}$. Since the announcement information is available to everyone, Eve obtains a copy (denoted by $E_{\widetilde{A}}$) of $\widetilde{A}$ and a copy (denoted by $E_{\widetilde{B}}$) of $\widetilde{B}$. The coherent version of copying is represented by the controlled NOT operation. Also, Alice and Bob each have a copy of the other party's announcement register, denoted by $\widetilde{B}_{\text{copy}}$ and $\widetilde{A}_{\text{copy}}$, respectively. For convenience of writing, we use $\widetilde{A}$ and $\widetilde{B}$ to refer to both Alice's and Bob's copies of $\widetilde{A}$ and $\widetilde{B}$. As we see later, we can actually combine the register $\widetilde{A}$ with $\widetilde{A}_{\text{copy}}$ and combine $\widetilde{B}$ with $\widetilde{B}_{\text{copy}}$ in the key rate calculation. In this diagram, the state after the announcement and measurement step is $\rho^{(1)}_{[A][B]EE_{\widetilde{A}}E_{\widetilde{B}}} =(W_A \otimes W_B) \rho_{ABE} (W_A \otimes W_B)^{\dagger}$, where for ease of writing, we reorder the registers and use a shorthand notation for collections of registers: $[A]$ for registers $A\widetilde{A}\overline{A}$ and $[B]$ for registers $B\widetilde{B}\overline{B}$.

The sifting step partitions the set of announcement events $S^A \times S^B$ as $S^A \times S^B = \mathbf{K} \cup \mathbf{D} $, where $\mathbf{K}$ is the set of announcement events to be kept and $\mathbf{D}$ is the set of announcements to be discarded. The sifting isometry (denoted by $V_S$) introduces a register $S$ to store the result of sifting (``keep'' or ``discard'') and performs a unitary operator $U_{\text{S}}$ on the local copies of registers $\widetilde{A}$ and $\widetilde{B}$ to compute the sifting decision. In a common scenario, each party can implement this unitary $U_{\text{S}}$ from the description of a protocol. If it is not from the protocol description and additional classical communication is needed, then, after a party implements this unitary operation, other parties obtain a copy of the register $S$. For simplicity, we use $S$ to refer to both Alice's and Bob's copies of this register. In the diagram, the state after this sifting step is $\rho^{(2)}_{[A][B]S[E]} = V_S \rho^{(1)}_{[A][B]EE_{\widetilde{A}}E_{\widetilde{B}}}V_S^{\dagger},$ where we use a shorthand notation $[E]$ to refer to Eve's collection of registers $EE_{\widetilde{A}}E_{\widetilde{B}}E_S.$

The key map isometry $V_K$ introduces a register $R$ and applies a local unitary $U_{\text{K}}$ to compute the key map $g$ and to store the result in the register $R$. This key map $g$ takes the announcement $(a, b) \in S^A \times S^B$ and Alice's measurement outcome $f_a(\alpha)$ in the case of direct reconciliation or Bob's measurement outcome $f_b(\beta)$ in the case of reverse reconciliation as inputs and outputs a value in $\{0,1,\dots, N-1\}$ where $N$ is the number of key symbols. For the purpose of derivation, we include an additional key symbol $\perp$ to this set and map all discarded events to it. We see later that we can eventually remove the symbol $\perp$ from the set of key symbols. In the diagram, the state after the key map state is $\rho^{(3)}_{R[A][B]S[E]} = V_K \rho^{(2)}_{[A][B]S[E]} V_K^{\dagger}$.

To give explicit expressions for these isometries $W_A$, $W_B$, $V_S$, and $V_K$, we first define $K_a^A$ and $K_b^B$ as
 \begin{aeq}\label{eq:appA_kraus_op}
 K_a^A &= \sum_{\alpha \in \Omega^A} \sqrt{P^A_{[a,f_a(\alpha)]}} \otimes \ket{a}_{\widetilde{A}} \otimes \ket{\alpha}_{\overline{A}}, \\
 K_b^B &=  \sum_{\beta \in \Omega^B} \sqrt{P^B_{[b,f_b(\beta)]}} \otimes \ket{b}_{\widetilde{B}} \otimes \ket{\beta}_{\overline{B}},\\
 \end{aeq}where $\{\ket{a}: a \in S^A\}$ and $\{\ket{b}: b \in S^B\}$ are orthonormal bases for registers $\widetilde{A}$ ($E_{\widetilde{A}}$) and $\widetilde{B}$ ($E_{\widetilde{B}}$) and $\{\ket{\alpha}: \alpha \in \Omega^A\}$ and $\{\ket{\beta}: \beta \in \Omega^B\}$ are orthonormal bases for registers $\overline{A}$ and $\overline{B}$, respectively. (Note that $\ket{\alpha}$ here is not a coherent state discussed in the main text.) We remark that $K_a^A$ and $K_b^B$ are the same as defined in Eqs. (40) and (41) of Ref. \cite{Winick2018} if we write $f_a(\alpha)$ as $\alpha_a$ and $f_b(\beta)$ as $\beta_b$. Then $W_A$, $W_B$ and $V_S$ are defined, respectively, as
\begin{aeq}\label{eq:isometry_announcement}
W_A &= \sum_{a \in S^A} K_a^A \otimes \ket{a}_{E_{\widetilde{A}}},\\
W_B  &=\sum_{b \in S^B} K_b^B \otimes  \ket{b}_{E_{\widetilde{B}}},\\
V_S &= \Pi \otimes \ket{\mathbf{K}}_S   \otimes \ket{\mathbf{K}}_{E_S} + (\mathds{1}_{\widetilde{A}\widetilde{B}}  - \Pi) \otimes \ket{\mathbf{D}}_S  \otimes \ket{\mathbf{D}}_{E_S},\\
\end{aeq}where $\Pi = \sum_{(a, b) \in \mathbf{K}} \dyad{a}{a}_{\widetilde{A}} \otimes \dyad{b}{b}_{\widetilde{B}}$ and $\{\ket{\mathbf{K}}, \ket{\mathbf{D}}\}$ is an orthonormal basis for the register $S$ ($E_S$). To write out the key map isometry $V_K$, we take the reverse reconciliation schemes as an example, and it is straightforward to write out $V_K$ in the case of direct reconciliation schemes by using Alice's measurement outcome $f_a(\alpha)$ instead of Bob's outcome $f_b(\beta)$. We first define an (partial) isometry $V$ on the subspace that $\Pi$ projects onto and then write out $V_K$: 
\begin{aeq}
V &= \sum_{\substack{(a,b) \in \mathbf{K}\\ \beta \in \Omega^B}}  \ket{g(a, b, f_b(\beta))}_R \otimes \dyad{a}{a}_{\widetilde{A}} \otimes \dyad{b}{b}_{\widetilde{B}} \otimes \dyad{\beta}{\beta}_{\overline{B}}, \\V_K &= V \Pi + \ket{\perp}_R \otimes (\mathds{1}_{\widetilde{A}\widetilde{B}}-\Pi).
\end{aeq}

We remark that the final state $\rho^{(3)}_{R[A][B]S[E]}=V_K V_S (W_A \otimes W_B) \rho_{ABE}  (W_A \otimes W_B)^{\dagger} V_S^{\dagger} V_K^{\dagger}$  is a pure state since $\rho_{ABE}$ is a pure state, and we apply only isometries to it.

\subsection{Removing the dependence on Eve's registers}

To access the key information, we use the projective measurement $\{Z_j=\dyad{j}{j}_R: j \in \{0,1,\dots, N-1, \perp\}\}$. Since the final state $\rho^{(3)}_{R[A][B]S[E]}$ is pure, we apply Theorem 1 of Ref. \cite{Coles2012} to rewrite conditional entropy $H(\mathbf{Z}|[E])$ as 
 \begin{aeq}\label{eq:conversion}
  H(\mathbf{Z}|[E]) = D(\rho^{(3)}_{R[A][B]S}||\sum\limits_{j}Z_j \rho^{(3)}_{R[A][B]S} Z_j),
 \end{aeq}where $ \rho^{(3)}_{R[A][B]S}=\Tr_{[E]}(\rho^{(3)}_{R[A][B]S[E]})$.
 We then define an announcement map $\mathcal{A}$ for an input state $\sigma$ as $ \mathcal{A} (\sigma) = \sum_{a \in S^A}\sum_{b \in S^B} (K_a^A \otimes K_b^B) \sigma (K_a^A \otimes K_b^B)^{\dagger}$ and rewrite $\rho^{(3)}_{R[A][B]S}$ as
\begin{aeq}
\rho^{(3)}_{R[A][B]S} =& \Tr_{[E]} (\rho^{(3)}_{R[A][B]S[E]}) \\ =&  p_{\text{pass}} \rho_{R[A][B]}^{\mathbf{K}} \otimes \dyad{\mathbf{K}}{\mathbf{K}}_S \\ &+  (1-p_{\text{pass}}) \rho_{R[A][B]}^{\mathbf{D}} \otimes \dyad{\mathbf{D}}{\mathbf{D}}_S,
\end{aeq}where $p_{\text{pass}}= \Tr(V \Pi \mathcal{A}(\rho_{AB}) \Pi V^{\dagger}) =\Tr(\mathcal{A}(\rho_{AB}) \Pi)$ is the same sifting probability defined in the main text and 
\begin{aeq}
\rho_{R[A][B]}^{\mathbf{K}} &= \frac{V \Pi \mathcal{A}(\rho_{AB}) \Pi V^{\dagger}}{ p_{\text{pass}}}, \\
\rho_{R[A][B]}^{\mathbf{D}} &= \frac{\dyad{\perp}{\perp}_R \otimes  (\mathds{1}_{\widetilde{A}\widetilde{B}}-\Pi) \mathcal{A}(\rho_{AB})(\mathds{1}_{\widetilde{A}\widetilde{B}}-\Pi)}{1-p_{\text{pass}}}.
\end{aeq}

To show that the symbol $\perp$ has no contribution to the key rate, we use the following lemma (see Ref. \cite{Wilde2013}).
\begin{lemma}\label{lemma:QCstate}
For quantum-classical states $\rho_{QX}$ and $\sigma_{QX}$ defined as $\rho_{QX} = \sum_x p(x) \rho_Q^x \otimes \dyad{x}{x}_X, \; \; \; \sigma_{QX} = \sum_x q(x) \sigma_Q^x \otimes \dyad{x}{x}_X,$ where $p$ and $q$ are probability distributions over a finite alphabet $\mathcal{X}$ and $\rho_Q^x$ and $\sigma_Q^x$ are density operators for all $x \in \mathcal{X}$, the quantum relative entropy is
$D(\rho_{QX} || \sigma_{QX}) = \sum_x p(x) D(\rho_Q^x || \sigma_Q^x) + D(p || q)$.
\end{lemma}

Applying the lemma to the state $\rho^{(3)}_{R[A][B]S}$ with the classical register $S$ gives us \begin{aeq}\label{eq:relative_entropy_simplified}
& \; \; \; D\big(\rho^{(3)}_{R[A][B]S} || \mathcal{Z}(\rho^{(3)}_{R[A][B]S})\big) \\&= p_{\text{pass}} D\big(\rho_{R[A][B]}^{\mathbf{K}} || \mathcal{Z}(\rho_{R[A][B]}^{\mathbf{K}})\big)  \\ & \; \; \; + (1-p_{\text{pass}}) D(\rho_{R[A][B]}^{\mathbf{D}} ||\rho_{R[A][B]}^{\mathbf{D}} ) \\
&= p_{\text{pass}} D\big(\rho_{R[A][B]}^{\mathbf{K}} || \mathcal{Z}(\rho_{R[A][B]}^{\mathbf{K}})\big) \\
&= D\big(\mathcal{G}(\rho_{AB}) || \mathcal{Z}[\mathcal{G}(\rho_{AB})]\big),
\end{aeq}where we define $\mathcal{G}(\rho_{AB}) = V \Pi \mathcal{A}(\rho_{AB}) \Pi V^{\dagger}$, which is the same as in Ref. \cite{Winick2018}.

Finally, we remark that, on the subspace where $\Pi$ projects, the symbol $\perp$ does not show up anymore. Thus, we can modify $\{Z_j\}$ to remove the symbol $\perp$ in the end. This modification gives back to the definition of $\mathcal{Z}$ shown in Ref. \cite{Winick2018}.

\subsection{Simplifying the postprocessing map}
We now provide several remarks to explain how we can simplify the map $\mathcal{G}$ while making sure that such a simplification does not change our calculated key rates. Our discussion here takes the reverse reconciliation schemes as an example. It is straightforward to adapt the arguments to the direct reconciliation schemes.

We first make a remark about the registers $\widetilde{A}_{\text{copy}}$ and $\widetilde{B}_{\text{copy}}$ that are hidden in our notation $\widetilde{A}$ and $\widetilde{B}$. After tracing out Eve's registers $E_{\widetilde{A}}$ and $E_{\widetilde{B}}$, Alice's register $\widetilde{A}$ and Bob's copy $\widetilde{A}_{\text{copy}}$ are both classical registers, and, likewise,  Bob's register $\widetilde{B}$ and Alice's copy $\widetilde{B}_{\text{copy}}$ are classical. Since each of the sifting and key map steps is done locally via a controlled unitary whose target is the register $S$ or $R$ alone, we can pull out two copies of registers $\widetilde{A}$ and $\widetilde{B}$ to write the final state in the form of the quantum-classical state to which the previous lemma applies. If we look at the block diagonal structure of $\mathcal{G}(\rho_{AB})$ with respect to two copies of the register $\widetilde{A}$, we see directly that the state with a single copy of the register $\widetilde{A}$ is just embedded in a larger space with two copies of the register $\widetilde{A}$. This result means the eigenvalues of the state are unaffected by removing one copy of the register $\widetilde{A}$. A similar argument works for two copies of $\widetilde{B}$. Moreover, from the previous lemma, we see immediately that we can calculate the key rate from individual announcements if we write the key map $g$ as $g[a, b, f_b(\beta)] =: g_{ab} (\beta)$ for a collection of functions $g_{ab}$, one for each $(a, b) \in \mathbf{K}$. In this case, for each $(a, b) \in \mathbf{K}$, we define an isometry $V_{ab} = \sum_{\beta \in \Omega^B}  \ket{g_{ab}(\beta)}_R   \otimes \dyad{\beta}{\beta}_{\overline{B}}$ and a completely positive map $\mathcal{G}_{ab}$ for an input state $\sigma$ as $\mathcal{G}_{ab} (\sigma) = V_{ab} (\widetilde{K}_a^A \otimes \widetilde{K}_b^B)\sigma (\widetilde{K}_a^A \otimes \widetilde{K}_b^B)^{\dagger} V_{ab}^{\dagger}$, where we define $\widetilde{K}_a^A$ such that $K_a^A = \widetilde{K}_a^A \otimes \ket{a}_{\widetilde{A}}$ and  $\widetilde{K}_b^B$ such that $K_b^B = \widetilde{K}_b^B \otimes \ket{b}_{\widetilde{B}}$. Then, 
\begin{aeq}\label{eq:appA_objective_QCStructure}
 &D\big(\mathcal{G}(\rho_{AB})||\mathcal{Z}[\mathcal{G}(\rho_{AB})]\big) \\
 =&\sum_{(a,b) \in \mathbf{K}} D\big(\mathcal{G}_{ab}(\rho_{AB}) || \mathcal{Z}[\mathcal{G}_{ab}(\rho_{AB})]\big).
\end{aeq}

Besides the lemma, our objective function has another important property. Since the quantum relative entropy is invariant under an isometry, if an isometry can commute with $\mathcal{G}$ and $\mathcal{Z}$ maps, then our objective function is also invariant under this isometry. In other words, we can add or remove an isometry $W$ (that acts only on Alice's and Bob's registers) in the final expression of Eq. (\ref{eq:relative_entropy_simplified}) if $W$ commutes with $\mathcal{G}$ and $\mathcal{Z}$ maps.

From this property of our objective function, if those functions $g_{ab}$'s are the identity function, then we see that each isometry $V_{ab}$ simply copies the register $\overline{B}$ and stores this copy to the register $R$. Adding this copy is a local isometry and renaming the register $\overline{B}$ by the name $R$ is a unitary. Thus, we can combine the registers $\overline{B}$ and $R$ and retain the name of $R$. 

\begin{figure}
\includegraphics[width =0.9\linewidth]{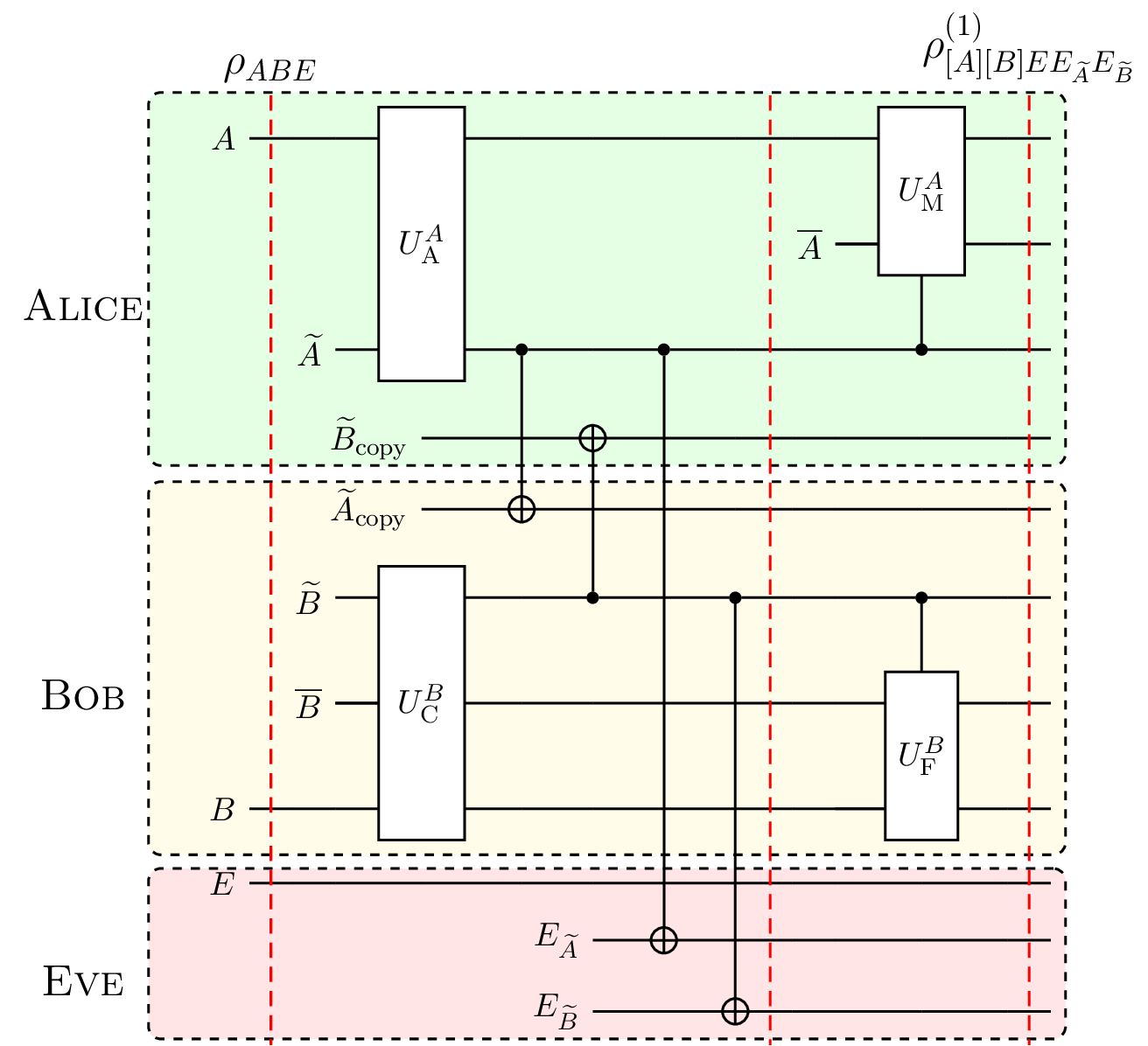}
\caption{\label{fig:pp_announcement} An alternative description of the announcement step for the reverse reconciliation schemes. This step can be decomposed into two steps. At the first step, Alice performs only a coarse-grained measurement with a unitary $U^A_{\text{A}}$ to obtain announcement results, and Bob also performs a coarse-grained measurement with a unitary $U^B_{\text{C}}$ to obtain announcement outcomes and coarse-grained measurement information. At the second step, they choose to perform optional refined measurements ($U^A_{\text{M}}$ and $U^B_{\text{F}}$) conditioned on the announcements (and previous coarse-grained measurement information for Bob). They can postpone the refined measurements after giving Eve announcement results and in some cases choose not to perform the refined measurements.}
\end{figure}

Also, based on this property of our objective function, we now discuss when we can omit the appearance of the register $\overline{A}$. As depicted in Fig. \ref{fig:pp_announcement}, the announcement and measurement step can be decomposed into two steps. First, Alice performs a coarse-grained measurement  (with associated unitary $U^A_{\text{A}}$ in this figure) to make announcements. Second, conditioned on her own announcement result, Alice performs a refined measurement (with a controlled unitary $U^A_{\text{M}}$ in the figure) if needed to obtain the fine-grained measurement outcomes. As the refined measurement is done in Alice's local registers, to which Eve has no access, Alice's refined measurement can be described by a local isometry. This local isometry can be performed after Eve obtains a copy of the announcement registers. In the reverse reconciliation scheme, since the key map isometry $V$ does not depend on the register $\overline{A}$, the isometry that describes Alice's refined measurement then commutes with $\mathcal{G}$ and $\mathcal{Z}$ maps. This result means that, from the key rate calculation perspective, we can drop this isometry due to the property of our objective function. More precisely, if we define $P_a^A = \sum_{\alpha \in \Omega^A} P^A_{[a,f_a(\alpha)]}$ for each $a \in S^A$ and define $K_a^{A'} = \sqrt{P_a^A} \otimes \ket{a}_{\widetilde{A}}$, then we can replace the map $\mathcal{G}$ by the map $\mathcal{G}'$ defined as 
\begin{aeq}
\mathcal{G}'(\sigma) =V \Pi \Big(\sum_{\substack{a \in S^A \\ b \in S^B}}  (K_a^{A'} \otimes K_b^B) \sigma (K_a^{A'} \otimes K_b^B)^{\dagger}\Big) \Pi V^{\dagger}
\end{aeq}for an input state $\sigma$. (A similar replacement can be done for $\mathcal{G}_{ab}$.) A similar argument can be applied to the direct reconciliation schemes by interchanging the roles of Alice's and Bob's registers to show that we can omit the register $\overline{B}$ for direct reconciliation.

Along the same line of argument, we remark that the refined measurement conditioned on the announcements can be a coarse-grained instead of the fine-grained measurement if the key map uses only the coarse-grained information. This process is also described in Fig. \ref{fig:pp_announcement}. Bob first applies an isometry (with associated unitary $U^B_{\text{C}}$ in the figure) to implement the coarse-grained measurement which gives the same coarse-grained information needed for the key map $g$ and then applies an additional isometry (with associated unitary $U^B_{\text{F}}$ in the figure) to obtain fine-grained measurement outcomes. As the sifting step depends only on the announcement, we can move Bob's refined measurement after the sifting step (not shown in this figure). Since the key map uses only the coarse-grained information, the key map isometry effectively undoes the unitary $U^B_{\text{F}}$. Therefore, we can take the POVM related to the coarse-grained measurement when we write out Kraus operators $K^B_b$ in Eq. (\ref{eq:appA_kraus_op}).

Finally, we explain how we derive the Kraus operators shown in Eqs. (\ref{eq:kraus_homo}) and (\ref{eq:kraus_het}). First, since we consider the reverse reconciliation schemes, we can omit the measurement outcome register $\overline{A}$. Second, since the key map of each protocol only uses the coarse-grained measurement outcomes, instead of using Bob's fine-grained POVM corresponding to homodyne or heterodyne measurements, we use the coarse-grained POVM ($\{I_0, I_1, \mathds{1} -I_0 -I_1\}$ for protocol 1 and $\{R_0, R_1, R_2, R_3, \mathds{1} -\sum_{j=0}^3 R_j\}$ for protocol 2). Since the set $\mathbf{K}$ contains only one element, we are left with only one term in the summation of Eq. (\ref{eq:appA_objective_QCStructure}) after removing registers $\widetilde{A}$ and $\widetilde{B}$. Finally, the key map in this case is the identity map. Thus, we combine registers $R$ and $\overline{B}$ and retain the name $R$ for this combined register.

\section{OPERATORS WITH THE PHOTON-NUMBER CUTOFF}\label{app:truncation}
Let $\mathcal{N}$ denote the photon-number basis up to $N_c$, that is, $\mathcal{N} = \{\ket{0}, \dots, \ket{N_c}\}$. In this work, we impose a photon-number cutoff assumption, that is, $\rho_{AB} = (\mathds{1}_A \otimes \Pi_{N_c}) \rho_{AB}  (\mathds{1}_A \otimes \Pi_{N_c})$ with the projection $\Pi_{N_c}$ onto the subspace spanned by the basis $\mathcal{N}$. Since Alice's system is irrelevant for our discussion here, we focus on the conditional states $\rho_B^x$ in the following discussion. For any operator $\hat{O}$ acting on Bob's system, we observe that
\begin{aeq}\label{eq:truncation}
\Tr[\rho_{B}^x \hat{O}] &= \Tr[\Pi_{N_c}\rho_{B}^x \Pi_{N_c} \hat{O}]  \\
& =  \Tr[(\Pi_{N_c}\rho_{B}^x \Pi_{N_c}) (\Pi_{N_c} \hat{O}\Pi_{N_c}) ]. 
\end{aeq}This observation allows us to define the truncated version of the operator $\hat{O}$ by $\Pi_{N_c} \hat{O}\Pi_{N_c}$. In our optimization problem [see Eq. (\ref{eq:optimization})], the relevant operators are of the forms $\Pi_{N_c}\rho_{B}^x \Pi_{N_c}$ and $\Pi_{N_c} \hat{O}\Pi_{N_c}$, which have finite-dimensional matrix representations. Specifically, we can find a matrix representation of $\hat{O}$ in the basis $\mathcal{N}$. We start by writing out the annihilation operator $\hat{a}$ in this basis, and then the creation operator $\hat{a}^{\dagger}$ is just its conjugate transpose. Consequently, other relevant operators $\hat{q}, \hat{p}$, $\hat{n}$, and $\hat{d}$ can be written directly following from their definitions in terms of $\hat{a}$ and $\hat{a}^{\dagger}$. In this basis,
\begin{aeq}
\Pi_{N_c} \hat{a}\Pi_{N_c} = \begin{pmatrix}
0 & 1& 0 & 0 & \cdots  & 0\\
0 & 0 & \sqrt{2} & 0&  \cdots & 0\\
\vdots &&\ddots   & &&  \vdots\\
 0 &&  \cdots  & & 0&  \sqrt{N_c} \\
 0 &&  \cdots  & & 0&  0 
\end{pmatrix}.
\end{aeq}

It is not difficult to write out the interval operators $I_0$ and $I_1$ and region operators $R_0, R_1, R_2$, and $R_3$ in this basis. To do so, we use the overlap $\bra{q}\ket{n}$ between a quadrature eigenstate $\ket{q}$ and a photon-number state $\ket{n}$ and the overlap $\bra{\gamma e^{i\theta}}\ket{n}$ between a coherent state $\ket{\gamma e^{i\theta}}$ and a photon-number state $\ket{n}$. With our definition of quadrature operators in Eq. (\ref{eq:quadrature_ops}), the overlaps $\bra{q}\ket{n}$ and $\bra{\gamma e^{i\theta}}\ket{n}$ read \cite{Barnett2002}
\begin{aeq}
\bra{q}\ket{n} &= \frac{1}{\sqrt{\pi^{\frac{1}{2}}{2^{n}(n!)}}} \exp(-\frac{q^2}{2}) H_n(q),\\
\bra{\gamma e^{i\theta}}\ket{n} &= e^{-\frac{\gamma^2}{2}} \frac{\gamma^n e^{-i n\theta}}{\sqrt{n!}},
\end{aeq}respectively, where $H_n$ is the Hermite polynomial of the order of $n$. We then perform the relevant integrals to obtain a finite-dimensional matrix representation in this basis.

Finally, in the expression of the Kraus operators shown in Eqs. (\ref{eq:kraus_homo}) and (\ref{eq:kraus_het}), we need to take the square root of each of the interval operators $I_0$ and $I_1$ or region operators $R_0, R_1, R_2$, and $R_3$. A caution about the ordering of truncation and square root is needed. For example, even though the interval operators are projective on the entire infinite-dimensional space such that the square root of each operator is identical to itself, the truncated version of each interval operator is no longer projective in the finite-dimensional subspace spanned by the basis $\mathcal{N}$. We now explain the proper way to handle this issue. With the photon-number cutoff assumption $\rho = \Pi_{N_c} \rho \Pi_{N_c}$, we see from Eq. (\ref{eq:truncation}) that, for a POVM element $F$ on the infinite-dimensional space, the corresponding POVM element on this finite-dimensional subspace becomes $\Pi_{N_c} F \Pi_{N_c}$. As we know from Appendix \ref{app:postprocessing}, the purpose of taking the square root of a POVM element is to realize this POVM measurement in an isometric fashion. Since the relevant POVM element on this finite-dimensional subspace is $\Pi_{N_c} F \Pi_{N_c}$, we need to take the square root of $\Pi_{N_c} F \Pi_{N_c}$. Therefore, we first take the truncation and then take the square root.

\section{EVALUATION OF LOSS-ONLY KEY RATES}\label{app:lossonly}
We discuss how to evaluate the Devetak-Winter formula in the loss-only scenario in the absence of postselection. When Alice sends $\ket{\alpha_x}_{A'}$ to Bob, in the absence of noise, Bob can verify that he receives a pure coherent state via homodyne or heterodyne detections. In the case of homodyne detection, Bob can verify that the received state is a minimum uncertainty state with the same variance for both quadratures, which implies it is a pure coherent state. In the case of heterodyne detection, Bob performs a tomography to verify that the received state is a pure coherent state. In particular, if Bob verifies his state to be an attenuated coherent state $\ket{\sqrt{\eta}\alpha_x}$, it is shown \cite{Heid2006} that Eve's optimal attack is the generalized beam-splitting attack for this pure-loss channel. Thus, the state shared by Bob and Eve becomes $\ket{\sqrt{\eta}\alpha_x}_B \ket{\sqrt{1-\eta}\alpha_x}_E$ after this channel. Because of the product state structure of Bob and Eve's joint state, Bob's measurement outcome does not influence Eve's state. Therefore, conditioned on the value $x$ of Alice's string $\mathbf{X}$ and the value $z$ in Bob's raw key $\mathbf{Z}$, Eve's conditional state $\ket{\epsilon_{x,z}}$ is
\begin{aeq}
\ket{\epsilon_{x,z}} =  \ket{\sqrt{1-\eta}\alpha_x} := \ket{\epsilon_x},
\end{aeq}which is independent from $z$ and, thus, we call it $\ket{\epsilon_x}$ for simplicity.
\subsection{Protocol 1}
The procedure outlined here is similar to the calculation in Ref. \cite{Heid2006}. For protocol 1, since $\alpha_x \in \{\alpha, -\alpha\}$, Eve's conditional states $\ket{\epsilon_{x}}$ are either $\ket{\sqrt{1-\eta}\alpha}$ or $\ket{-\sqrt{1-\eta}\alpha}$, which span only a two-dimensional subspace. We can find a two-dimensional representation of $\ket{\epsilon_{x}}$ as
\begin{aeq}
\ket{\epsilon_{0}}&= \ket{\sqrt{1-\eta}\alpha} = c_0 \ket{e_0} + c_1 \ket{e_1}, \\
\ket{\epsilon_{1}}&=\ket{-\sqrt{1-\eta}\alpha} = c_0 \ket{e_0} - c_1 \ket{e_1},
\end{aeq}where $\ket{e_0}$ and $\ket{e_1}$ are defined as
 \begin{equation}
\begin{aligned}
\ket{e_0} &= \frac{1}{\sqrt{\cosh[(1-\eta)\alpha^2]}}\sum_{n=0}^{\infty}\frac{(\sqrt{1-\eta}\alpha)^{2n}}{\sqrt{(2n)!}}\ket{2n},\\
\ket{e_1} &= \frac{1}{\sqrt{\sinh[(1-\eta)\alpha^2]}}\sum_{n=0}^{\infty}\frac{(\sqrt{1-\eta}\alpha)^{2n+1}}{\sqrt{(2n+1)!}}\ket{2n+1},
\end{aligned}
\end{equation}
respectively, $ c_0 = e^{-\frac{(1-\eta)\alpha^2}{2}}\sqrt{\cosh[(1-\eta)\alpha^2]}$ and $c_1 = e^{-\frac{(1-\eta)\alpha^2}{2}}\sqrt{\sinh[(1-\eta)\alpha^2]}.$

We now directly evaluate the Devetak-Winter formula
\begin{aeq}
R^{\infty} = \beta I(\mathbf{X}; \mathbf{Z}) - \chi( \mathbf{Z}{\,:\,}E).
\end{aeq}We obtain $I(\mathbf{X}; \mathbf{Z})$ by a calculation similar to Eq. (\ref{eq:delta_ec}). We can directly calculate $\chi(\mathbf{Z}{\,:\,}E)$ via
\begin{aeq}\label{eq:holevo}
\chi(\mathbf{Z}{\,:\,}E) = H(\overline{\rho}_{E}) - \sum_{j=0}^{1} P(z = j) H(\rho_{E,j}),
\end{aeq}where $H(\sigma) = -\Tr(\sigma \log_2\sigma)$ is the von Neumann entropy and the relevant states are
\begin{aeq}\label{eq:conditional_states}
\rho_{E,j} &= \sum_{i=0}^1 \frac{P(x=i, z = j)}{ P( z = j)} \dyad{\epsilon_i}{\epsilon_i}, \\
\overline{\rho}_E& =\sum_{j=0}^{1} P(z = j) \rho_{E,j},
\end{aeq}where $P(x, z)$ is the joint probability distribution of $x$ and $z$ and $P(z)$ is the marginal probability distribution of $z$. Each of the relevant Eve's states $\ket{\epsilon_x}$ has a two-dimensional matrix representation in the basis $\{\ket{e_0}, \ket{e_1}\}$, and, thus, it is straightforward to directly evaluate the Devetak-Winter formula.

\subsection{Protocol 2}

For the protocol 2, since $\alpha_x \in \{\alpha, i\alpha, -\alpha, -i\alpha\}$, Eve's conditional states $\ket{\epsilon_{x}}$ are  $\ket{\sqrt{1-\eta}\alpha}$, $\ket{i\sqrt{1-\eta}\alpha}$,  $\ket{-\sqrt{1-\eta}\alpha}$, and $\ket{-i\sqrt{1-\eta}\alpha}$, which span only a four-dimensional subspace. Therefore, we can find an orthonormal basis $\{\ket{f_0},\ket{f_1},\ket{f_2},\ket{f_3}\}$ for this subspace similar to the basis $\{\ket{e_0},\ket{e_1}\}$ and find a four-dimensional matrix presentation for each of Eve's conditional states (see Ref. \cite{Dusek2000} for an explicit expression). All the procedures are similar to protocol 1 except that the summation indexes $i$ and $j$ now run from 0 to 3 instead of 0 to 1 in Eqs. (\ref{eq:holevo}) and (\ref{eq:conditional_states}). With a four-dimensional matrix representation of Eve's conditional states $\ket{\epsilon_{x}}$ in the basis $\{\ket{f_0},\ket{f_1},\ket{f_2},\ket{f_3}\}$, it is also straightforward to directly evaluate the Devetak-Winter formula.

\bibliographystyle{apsrev4-2}
\bibliography{DMCVQKD}

\end{document}